\documentclass[aps,prd,showpacs,twocolumn]{revtex4}
\usepackage{graphicx}
\usepackage{amssymb}
\usepackage{epsfig}
\usepackage{bm}
\usepackage{dcolumn}

\newcommand{\etal}{et al.}

\begin{document}

\title{Hints of the existence of Axion-Like-Particles from the gamma-ray spectra \\of cosmological sources}

\author{M. A. S\'anchez-Conde$^{1,3}$, D. Paneque$^{2}$, E. Bloom$^2$, F. Prada$^{1,5}$ and A. Dom\'inguez$^{1,4}$}
\affiliation{$^1$ Instituto de Astrof\'isica de Andaluc\'ia (CSIC), E-18008, Granada, Spain} 
\email{masc@iaa.es}
\affiliation{$^2$ Kavli Institute for Particle Astrophysics and Cosmology (KIPAC), SLAC National Accelerator Center, Sand Hill Road 2575, CA 94025, USA}
\email{dpaneque@slac.stanford.edu}
\affiliation{$^3$ Visiting student at SLAC National Accelerator Center}
\affiliation{$^4$ Departamento de F\'isica At\'omica, Molecular y Nuclear, Universidad de Sevilla, E-41012, Sevilla, Spain}
\affiliation{$^5$ Visiting research physicist at the Santa Cruz Institute for Particle Physics (SCIPP), University of California, Santa Cruz CA 95064, USA}
\date{\today}

\begin{abstract}
Axion Like Particles (ALPs) are predicted to couple with photons in the presence of magnetic fields. This effect may lead to a significant change in the observed spectra of gamma-ray sources such as AGNs. Here we carry out a detailed study that for the first time simultaneously considers in the same framework both the photon/axion mixing that takes place in the gamma-ray source and that one expected to occur in the intergalactic magnetic fields. An efficient photon/axion mixing in the source always means an attenuation in the photon flux, whereas the mixing in the intergalactic medium may result in a decrement and/or enhancement of the photon flux, depending on the distance of the source and the energy considered. Interestingly, we find that decreasing the value of the intergalactic magnetic field strength, which decreases the probability for photon/axion mixing, could result in an increase of the expected photon flux at Earth if the source is far enough. We also find a 30\% attenuation in the intensity spectrum of distant sources, which occurs at an energy that only depends on the properties of the ALPs and the intensity of the intergalactic magnetic field, and thus independent of the AGN source being observed. Moreover, we show that this mechanism can easily explain recent puzzles in the spectra of distant gamma-ray sources, like the possible detection of TeV photons from 3C~66A (a source located at z=0.444) by MAGIC and VERITAS, which should not happen according to conventional models of photon propagation over cosmological distances. Another puzzle is the recent published lower limit to the EBL intensity at 3.6 $\mu$m (which is almost twice larger as the previous one), which implies very hard spectra for some detected TeV gamma-ray sources located at z=0.1-0.2. The consequences that come from this work are testable with the current generation of gamma-ray instruments, namely Fermi (formerly known as GLAST) and imaging atmospheric Cherenkov telescopes like CANGAROO, HESS, MAGIC and VERITAS.

\end{abstract}


\maketitle

\section{Introduction}

The existence of axions is predicted by the  Peccei-Quinn mechanism, which is currently the most compelling explanation to solve the CP problem in QCD \cite{PQ}. Moreover, amongst all the valid candidates proposed to constitute a portion or the totality of the non-barionic cold dark matter content predicted to exist in the Universe, hypothetical non-thermal axions, or in a more generic way, Axion-Like Particles (ALPs), where the mass and the coupling constant are not related to each other, may represent a good option: they might exist in sufficient quantities to account for the estimated dark matter density and they might interact very weakly with the rest of the particles \cite{raffelt}.  There is an additional property of ALPs that makes them even more attractive and that could have important implications for its detectability, i.e. they can oscillate into photons and vice-versa in the presence of an electric or magnetic field \cite{dicus,sikivie}. This is analogous to that predicted to occur between neutrinos of different flavors, and a similar behavior is expected in the case of the recently proposed chameleons as well \cite{camaleon}. This characteristic is the main vehicle used at present to carry out an exhaustive search of ALPs by experiments like CAST \cite{cast}, PVLAS \cite{pvlas} and ADMX \cite{admx}.

The oscillation of photons to ALPs (and vice-versa) could have important implications for astronomical observations. This argument was first investigated in the optical band by Ref.~\cite{csaki02}, where authors proposed the existence of axions to be the cause of the observed supernova Ia dimming. In this context, the observed dimming might be explained as a result of an efficient photon to axion conversion instead of a cosmic acceleration (albeit this proposal was rejected some time later due to some chromatic problems, pointed out e.g. in Ref.~\cite{mirizzi06}). Photon/axion oscillations were also studied by the same authors in Ref.~\cite{csaki03} as an alternative explanation for those photons arriving Earth from very distant sources at energies above the GZK cutoff. 

Recently, it has been proposed that, if ALPs exist, they could distort the spectra of gamma-ray sources, such as Active Galactic Nuclei (AGNs) \cite{hooper,deangelis,hochmuth,simet} or galactic sources in the TeV range \cite{mirizzi07}, and that their effect may be detected by current gamma-ray experiments. In \cite{burrage}, for example, it is stated that also the scatter in AGN luminosity relations could be used to search for ALPs. Other astrophysical environments have been proposed in order to detect ALPs, such as the magnetic field of the Sun \cite{fairbairn}, pulsars \cite{dupays}, the galactic halo \cite{duffy} or GRBs and QSOs by carefully studying their polarized gamma-ray emissions \cite{rubbia,hutsemekers}. In particular, these predictions are very relevant for gamma-ray astronomy, where recent instrumentation developments in the last few years have increased the observational capabilities by more than one order of magnitude. On the ground, we have the new generation of Imaging Atmospheric Cherenkov Telescopes (IACTs) like MAGIC \cite{magic}, HESS \cite{hess}, VERITAS \cite{veritas} or CANGAROO-III \cite{cangaroo}, covering energies in the range 0.1-20 TeV. In space we have Fermi (previously called GLAST) \cite{glast}, in operation since Summer 2008 and covering energies in the range 0.02-300 GeV\footnote{In the space we also have a new gamma-ray instrument called AGILE \cite{agile}, yet the sensitivity is actually similar to that of EGRET}.

In this work we revisit the photon/axion mixing, for the first time handling under the same consistent framework the mixing that takes place inside or near the gamma-ray sources together with that one expected to occur in intergalactic magnetic field (IGMF). In the literature, both effects have been considered separately. Depending on the source dimension, magnetic field, ALP mass and coupling constant, both effects might produce significant spectral distortions, or one effect could be more important than the other. In any case, we believe that both effects could be relevant and hence need to be considered simultaneously. We neglect, however, the mixing that may happen inside the Milky Way due to galactic magnetic fields. At present, a concise modeling of this effect is still very dependent on the largely unknown morphology of the magnetic field in the galaxy. Furthermore, in the most idealistic/optimistic case, this effect would produce an enhancement of the photon flux arriving at Earth of
about 3\% of the initial photon flux emitted by the source \cite{simet}. This is in contrast with what we found for the IGMFs: although there is also little information on the strength and morphology of the IGMFs, the derived photon/axion mixing in this case we show to be crucial for a correct interpretation of the observed flux. It is worth mentioning that we will come to this conclusion using a conservative value of B=0.1~nG for the IGMF strength, well below the current upper limits of $\sim$1 nG. We also carry out a detailed analysis of the mixing when varying IGMF strength and source distance. We find results that differ from previously published ones, and we make predictions of effects that have not been noted in the literature so far.

At energies larger than 10 GeV, and especially above 100 GeV, it will be necessary to properly account for the Extragalactic Background Light (EBL) in our IGMF mixing calculations. The EBL introduces an attenuation in the photon flux due to $e^{-}e^{+}$ pair production that comes from the interaction of the gamma-ray source photons with infrared and optical-UV background photons \cite{ebl}. Amongst all the EBL models that exist in the literature, in this work we will make use of the Primack \cite{primack05} and Kneiske best-fit \cite{kneiske04} EBL models. They represent respectively one of the most transparent and one of the most opaque models for gamma-rays, but still within the limits imposed by the observations. The EBL model will play a crucial role in our formalism and results: as we will see, the more attenuating the EBL model considered, the more relevant the effect of photon/axion oscillations in the IGMF.

We also explore in this work the detection prospects for current gamma-ray instruments (Fermi and IACTs). We will show that the signatures of photon/axion oscillations may be observationally detectable provided light ALPs with masses smaller than a given value for typical values of the IGMF. In order to study the detection prospects, we will propose an observational strategy. We can anticipate here that the main challenge for our proposed formalism to be testable comes from the lack of knowledge of the intrinsic source spectrum and EBL density. However, we note that there is the possibility that we could be already detecting the first hints of axions with current experiments. In this context, the potential detection of TeV photons from very distant (z $\sim$ 0.4) sources \cite{acciari09,Albert2008_3c66a,Nesphor1998,Stepanyan2002}, or some works claiming energy spectral indices harder than 1.5 for relatively distant (z=0.1-0.2) AGNs \cite{Krennrich2008}, already put in a tight spot the conventional interpretation of the observed gamma-ray data. As we will show, both effects could be explained by oscillations of photon into light ALPs using realistic values for the involved parameters.

The work is organized as follows. In Section \ref{sec2} we describe in detail the photon/axion mixing in both the surroundings of gamma-ray sources and in the intergalactic medium (IGM). Section \ref{sec3} is devoted to present the results obtained when including both mixings under the same framework and after considering realistic parameters for well-known AGNs. In Section \ref{sec4} we present an observational strategy to search for ALPs using the most sensitive gamma-ray instruments, namely Fermi and IACTs like MAGIC or HESS. Finally, we give our conclusions in Section \ref{sec5}.

\section{The formalism} \label{sec2}

\begin{figure*}[!ht]
\centering
\includegraphics[height=5cm,width=14cm]{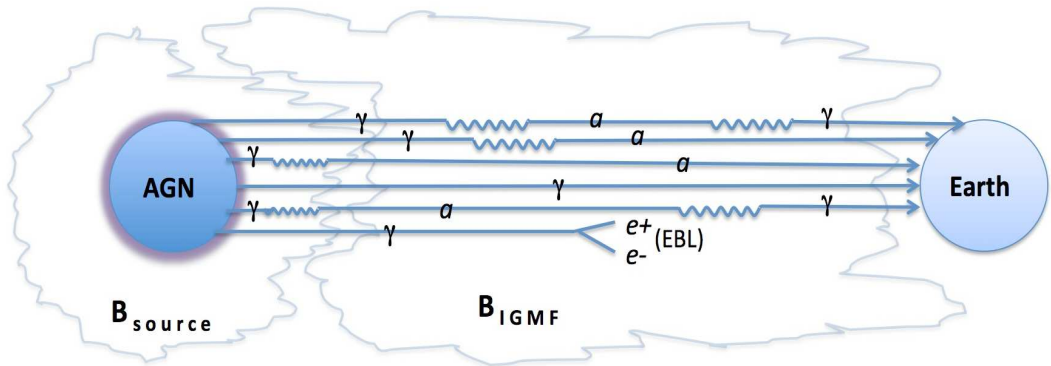}
\caption{\small{Sketch of the formalism used in this work, where both mixing inside the source and mixing in the IGMF are considered under the same consistent framework. Photon to axion oscillations (or vice-versa) are represented by a crooked line, while the symbols {\it $\gamma$} and {\it a} mean gamma-ray photons and axions respectively. This diagram collects the main physical scenarios that we might identify inside our formalism. Each of them are squematically represented by a line that goes from the source to the Earth.}}
\label{fig:sketch}
\end{figure*}

At present, the Peccei-Quinn mechanism remains as the most convincing solution to solve the CP violation of QCD. As early as in 1978, Weinberg \cite{weinberg} and Wilczek \cite{wilczek} realized independently that a consequence of this mechanism is the existence of a pseudo-scalar boson, the axion. One generic property of axions is a two-photon interaction of the form:

\begin{equation}
{\cal L}_{a \gamma} = -\frac{1}{4~M}~F_{\mu \nu}\overline{F}^{\mu \nu}a = 
\frac{1}{M}~{\bf E \cdot B}~a
\label{eq:lagrangiano}
\end{equation}

\noindent where $a$ is the axion field, $M$ is the inverse of the photon/axion coupling strength, $F$ is the electromagnetic field-strength tensor, $\overline{F}$ its dual, ${\bf E}$ the electric field, and ${\bf B}$ the magnetic field. The axion has the important feature that its mass $m_a$ and coupling constant are inversely related to each other. There are, however, other predicted states where this relation does not hold; such states are known as Axion Like Particles (ALPs). An important and intriguing consequence of Eq.~(\ref{eq:lagrangiano}) is that ALPs oscillate into photons and vice-versa in the presence of an electric or magnetic field. In fact this effect represents the keystone in ongoing ALP searches carried out by current experiments. 

In this work, we will make use of the photon/axion mixing as well, but this time by means of astrophysical magnetic fields. As already mentioned, we will account for the mixing that takes place inside or near the gamma-ray sources together with that one expected to occur in the IGMFs. We will do it under the same consistent framework. Furthermore, it is important to remark that it will be necessary to include the EBL in our formalism, in particular in the equations that describe the intergalactic mixing. Its main effect we should remember is an attenuation of the photon flux, especially at energies above 100 GeV. We show in Fig.~\ref{fig:sketch} a diagram that outlines our formalism. Very squematically, the diagram shows the travel of a photon from the source to the Earth in a scenario with axions. In the same Figure, we show the main physical cases that one could identify inside our formalism: mixing in both the source and the IGMF, mixing in only one of these environments, the effect of the EBL, axion to photon reconversions in the IGMF, etc. A quantitative description of the photon/axion mixing phenomenon in both the source and the IGMFs can be found in the next two subsections.

\subsection{Mixing inside and near the source} \label{sec_source}

It has been recently proposed that an efficient conversion from photons to ALPs (and vice-versa) could take place in or near some astrophysical objects that should host a strong magnetic field \cite{hooper}. 

Given a domain of length $s$, where there is a roughly constant magnetic field and plasma frequency, the probability of a photon of energy $E_{\gamma}$ to be converted into an ALP after traveling through it can be written as \cite{mirizzi07,hochmuth}:

\begin{equation}
P_0 = (\Delta_B~s)^2~\frac{\sin^2(\Delta_{osc}~s/2)}{(\Delta_{osc}~s/2)^2} 
\label{eq:P0}
\end{equation}

\noindent Here $\Delta_{osc}$ is the oscillation wave number: 
\begin{equation}
\Delta_{osc}^2 \simeq (\Delta_{CM}+\Delta_{pl}-\Delta_a)^2+4\Delta_B^2,
\label{eq:deltaosc}
\end{equation}

\noindent $\Delta_B$ that gives us an idea of how effective is the mixing, i.e.
\begin{equation}
\Delta_B = \frac{B_t}{2~M} \simeq 1.7 \times 10^{-21}~M_{11}~B_{mG}~cm^{-1},
\label{eq:deltaB}
\end{equation}
\noindent where $B_t$ the magnetic field component along the polarization vector of the photon and $M_{11}$ the inverse of the coupling constant.

$\Delta_{CM}$ is the vacuum Cotton-Mouton term, i.e.
\begin{eqnarray}
\Delta_{CM} &=& -\frac{\alpha}{45\pi}~\left(\frac{B_t}{B_{cr}}\right)^2E_{\gamma} \nonumber \\
&\simeq& -1.3 \times 10^{-21}~B^2_{mG}\left(\frac{E_{\gamma}}{TeV}\right)~cm^{-1}, 
\label{eq:deltaCM}
\end{eqnarray}
\noindent where $B_{cr}=m^2_e/e \simeq 4.41 \times 10^{13}$~G the critical magnetic field strength ($e$ is the electron charge).

$\Delta_{pl}$ is the plasma term:
\begin{equation}
\Delta_{pl} = \frac{w^2_{pl}}{2E} \simeq 3.5 \times 10^{-20}\left(\frac{n_e}{10^3cm^{-3}}\right)\left(\frac{TeV}{E_{\gamma}}\right)~cm^{-1},
\label{eq:deltapl}
\end{equation}
\noindent where $w_{pl}=\sqrt{4\pi\alpha n_e/m_e} = 0.37 \times 10^{-4} \mu eV \sqrt{n_e/cm^{-3}}$ the plasma frequency, $m_e$ the electron mass and $n_e$ the electron density.

Finally, $\Delta_a$ is the ALP mass term:
\begin{equation}
\Delta_{a} = \frac{m^2_{a}}{2E_{\gamma}} \simeq 2.5 \times 10^{-20}m^2_{a,\mu eV}\left(\frac{TeV}{E_{\gamma}}\right)~cm^{-1}.
\label{eq:deltaa}
\end{equation}

Note that in Eqs.(\ref{eq:deltaB}-\ref{eq:deltaa}) we have introduced the dimensionless quantities $B_{mG}=B/10^{-3}$ G, $M_{11}=M/10^{11}$ GeV and $m_{\mu eV}=m/10^{-6}$ eV.

Since we expect to have not only one coherence domain but several domains with magnetic fields different from zero and subsequently with a potential photon/axion mixing in each of them, we can derive a total conversion probability \cite{mirizzi07} as follows:

\begin{equation}
P_{\gamma \rightarrow a} \simeq \frac{1}{3}[1-\exp(-3NP_0/2)]
\label{eq:totalProb}
\end{equation}

\noindent where $P_0$ is given by Eq.(\ref{eq:P0}) and $N$ represents the number of domains. Note that in the limit where $N~P_0 \rightarrow \infty$, the total probability saturates to 1/3, i.e. one third of the photons will convert into ALPs.

It is useful here to rewrite Eq.~(\ref{eq:P0}) following Ref.~\cite{hooper}, i.e.

\begin{equation}\label{eq:prob2}
P_0 =\frac{1}{1+(E_{crit}/E_{\gamma})^2}~
\sin^2\left[\frac{B~s}{2~M}\sqrt{1+\left(\frac{E_{crit}}{E_{\gamma}}\right)^2}\right]
\end{equation}

\noindent so that we can define a characteristic energy, $E_{crit}$, given by:
\begin{equation} 
E_{crit} \equiv \frac{m^2~M}{2~B}
\label{eq:ecrit1}
\end{equation}

\noindent or in more convenient units:
\begin{equation}
E_{crit} (GeV) \equiv \frac{m^2_{\mu eV}~M_{11}}{0.4~B_G}
\label{eq:ecrit}
\end{equation}

\noindent where the subindices refer again to dimensionless quantities: $m_{\mu eV} \equiv m/ \mu eV$, $M_{11} \equiv M/10^{11}$ GeV and $B_G \equiv $ B/Gauss; $m$ is the effective ALP mass $m^2 \equiv |m_a^2-\omega_{pl}^2|$. Recent results from the CAST experiment \cite{cast} give a value of $M_{11} \geq 0.114$ for axion mass $m_a \leq 0.02$ eV. Although there are other limits derived with other methods or experiments, the CAST bound is the most general and stringent limit in the range $10^{-11}$ eV $\ll m_a \ll 10^{-2}$ eV.

At energies below $E_{crit}$ the conversion probability is small, which means that the mixing will be small. Therefore we must focus our detection efforts at energies above this $E_{crit}$, where the mixing is expected to be large ({\it strong mixing regime}). As pointed out in Ref.~\cite{hooper}, in the case of using typical parameters for an AGN in Eq.~(\ref{eq:ecrit}), $E_{crit}$ will lie in the GeV range given an ALP mass of the order of $\sim\mu$eV.

\begin{figure}[!h]
\centering
\includegraphics[height=6.5cm,width=8cm]{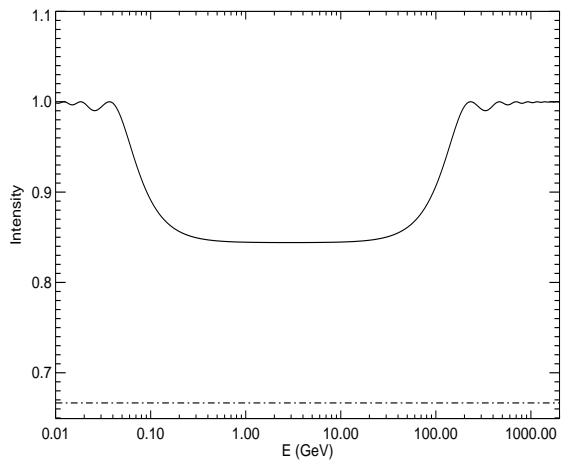}
\caption{\small{Example of photon/axion oscillations inside the source or vicinity, and its effect on the source intensity (solid line), which was normalized to 1 in the Figure.  We used the parameters given in Table~\ref{tab:3c279} to model the AGN source, but we adopted an ALP mass of 1 $\mu$eV. This gives $E_{crit}=0.19$ GeV. The dot-dashed line represents the maximum (theoretical) attenuation given by Eq.~(\ref{eq:totalProb}), and equal to 1/3.}} 
\label{fig:sourcemix}
\end{figure}

To illustrate how the photon/axion mixing inside the source works, we show in Figure~\ref{fig:sourcemix} an example for an AGN modeled by the parameters listed in Table~\ref{tab:3c279} (our fiducial model, see Section \ref{sec3}). The only difference is the use of an ALP mass of 1 $\mu$eV instead of the value that appear in that Table, so that we obtain a critical energy that lie in the GeV energy range; we get $E_{crit}=0.19$ GeV according to Eq.~(\ref{eq:ecrit}). Note that the main effect just above this critical energy is an attenuation in the total expected intensity of the source. However, note also that the attenuation starts to decrease at higher energies ($>$10 GeV) gradually. The reason for this behavior is the crucial role of the Cotton-Mouton term at those high energies, which makes the efficiency of the source mixing to decrease as the energy increases (see Eq.~(\ref{eq:deltaCM}) and how it affects to Eq.~(\ref{eq:deltaosc})). Indeed, the photon attenuation induced by the mixing in the source completely dissapears at energies above around 200 GeV in this particular example. On the other hand, one can see in Figure~\ref{fig:sourcemix} a sinusoidal behavior just below the critical energy as well as just below the energy at which the source mixing dissapears due to the Cotton-Mouton term. However, it must be noted that a) the oscillation effects are small; b) these oscillations only occur when using photons polarized in one direction while, in reality, the photon fluxes are expected to be rather non-polarized; and c) the above given expressions are approximations and actually only their asymptotic behavior should be taken as exact and well described by the formulae. Therefore, the chances of observing sinusoidally-varying energy spectra in astrophysical source, due to photon/axion oscillations, are essentially zero.

\subsection{Mixing in the IGMFs} \label{sec:intergalmix}

The strength of the Intergalactic Magnetic Fields (IGMFs) is expected to be many orders of magnitude weaker ($\sim$nG) than that of the source and its surroundings ($\sim$G). Consequently, as described by Eq.~(\ref{eq:ecrit}), the energy at which photon/axion oscillation occurs in the IGM is many orders of magnitude larger than that at which oscillation can occur in the source and its vicinity. Despite the low magnetic field {\bf B}, the photon/axion oscillation can take place due to the large distances, since the important quantity defining the probability for this conversion is the product B $\times$ s, as described by Eq~(\ref{eq:prob2}). Assuming B $\sim$0.1 nG (see below), and $M_{11}=0.114$ (coincident with the upper limit reported by CAST \cite{cast}), then the effect can be observationally detectable ($E_{crit}< 1$~TeV) only if the ALP mass is m$_a<6 \times 10^{-10}$ eV. If the axion mass m$_a$ was larger than this value, then the consequences of this oscillation could not be probed with the current generation of IACTs, that observe up to few tens of TeV \footnote{The next generation of IACTs (namely AGIS and CTA) aim for an order of magnitude of improvement at the highest energies, reaching few hundreds of TeV; but those instruments will not be in operation till 2013 or later.}. In our fiducial model (see Table \ref{tab:3c279}) we used m$_a=10^{-10}$ eV, which implies $E_{crit} = 28.5$ GeV.

It is important to stress that at energies larger than 10 GeV, and especially larger than 100 GeV, besides the oscillation to ALPs, the photons should also be affected by the diffuse radiation from the Extragalactic Background Light (EBL). The EBL introduces an attenuation in the photon flux due to $e^{-}e^{+}$ pair production that comes from the interaction of the gamma-ray source photons with infrared and optical-UV background photons for the energies under consideration \cite{ebl}. Therefore, it will be necessary to modify the above equations to properly account for the EBL in our calculations. These equations can be found in Ref.~\cite{csaki03}, where the photon/axion mixing in the IGMF was also studied, although for other purposes and a different energy range. We note that the same equations were also used in Ref.~\cite{deangelis} to study for the first time the photon/axion mixing in the presence of IGMFs for the same energy range that we are considering in this work. 

There is little information on the strength and morphology of the IGMFs. As for the morphology, several authors reported that space should be divided into several domains, each of them with a size for which the magnetic field is coherent. Different domains will have randomly changing directions of {\bf B} field of about the same strength \cite{Kronberg1994,Furlanetto2001}. The IGMF strength is constrained to be smaller than 1~nG \cite{grasso}, which is somewhat supported by the estimates of $\sim$0.3-0.9~nG  that can be inferred \cite{deangelisBfield} from recent observations of  the Pierre Auger Observatory \cite{auger}. On the other hand, there is controversy on the possibility of generating such a strong magnetic field. Detailed simulations yield IGMFs of the order of 0.01~nG so that they can later reproduce the measured ${\bf B}$ fields in nearby galaxy clusters \cite{dolag,sigl}. Given this controversy, we decided to use a mid-value of 0.1nG in our fiducial model (Table~\ref{tab:3c279}).  

\begin{figure*}[!ht]
  \centering
  \begin{minipage}[b]{0.32\textwidth}
    \centering
    \includegraphics[height=5cm,width=6cm]{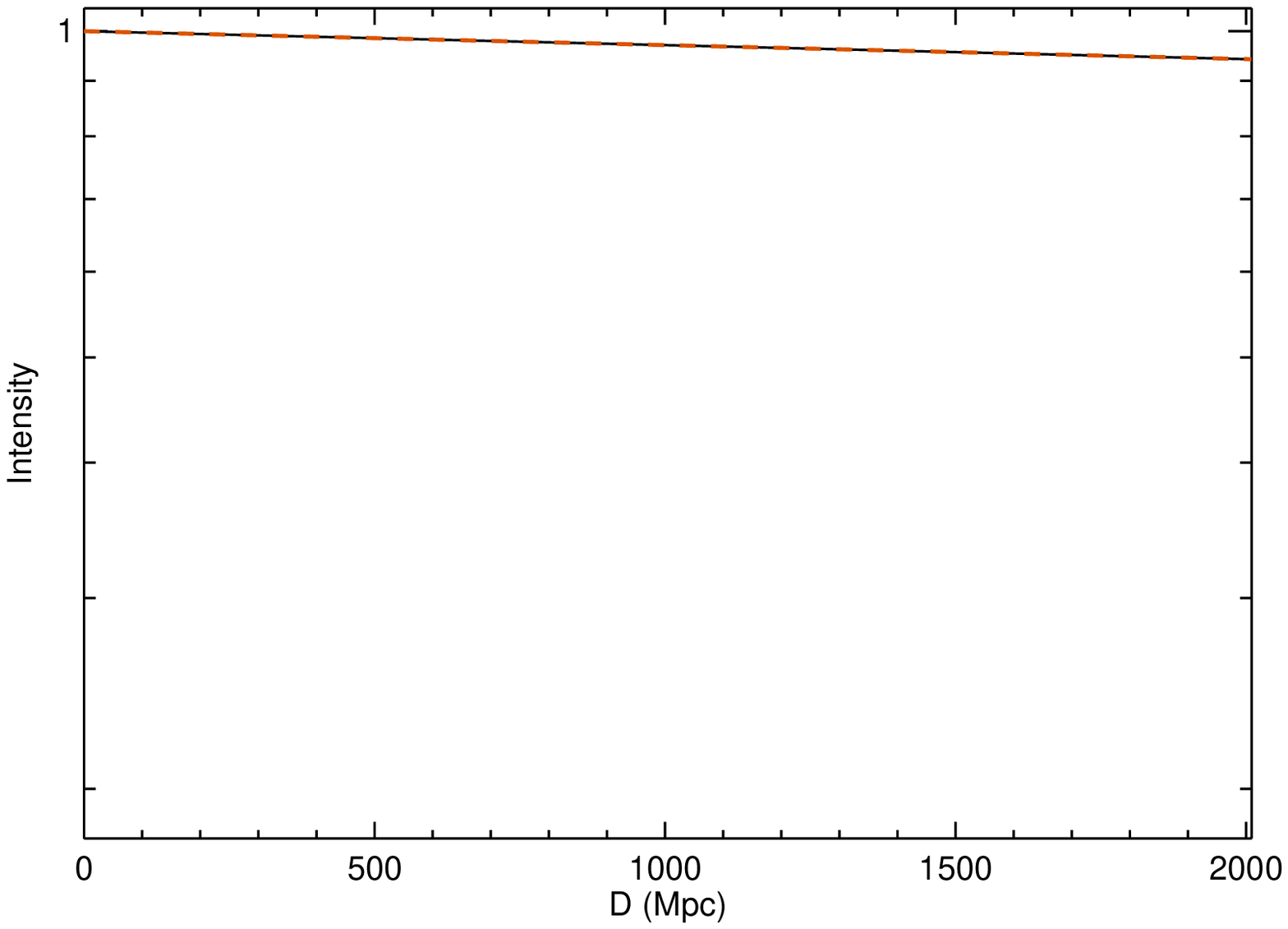}
  \end{minipage}
  \begin{minipage}[b]{0.32\textwidth}
    \centering
    \includegraphics[height=5cm,width=6cm]{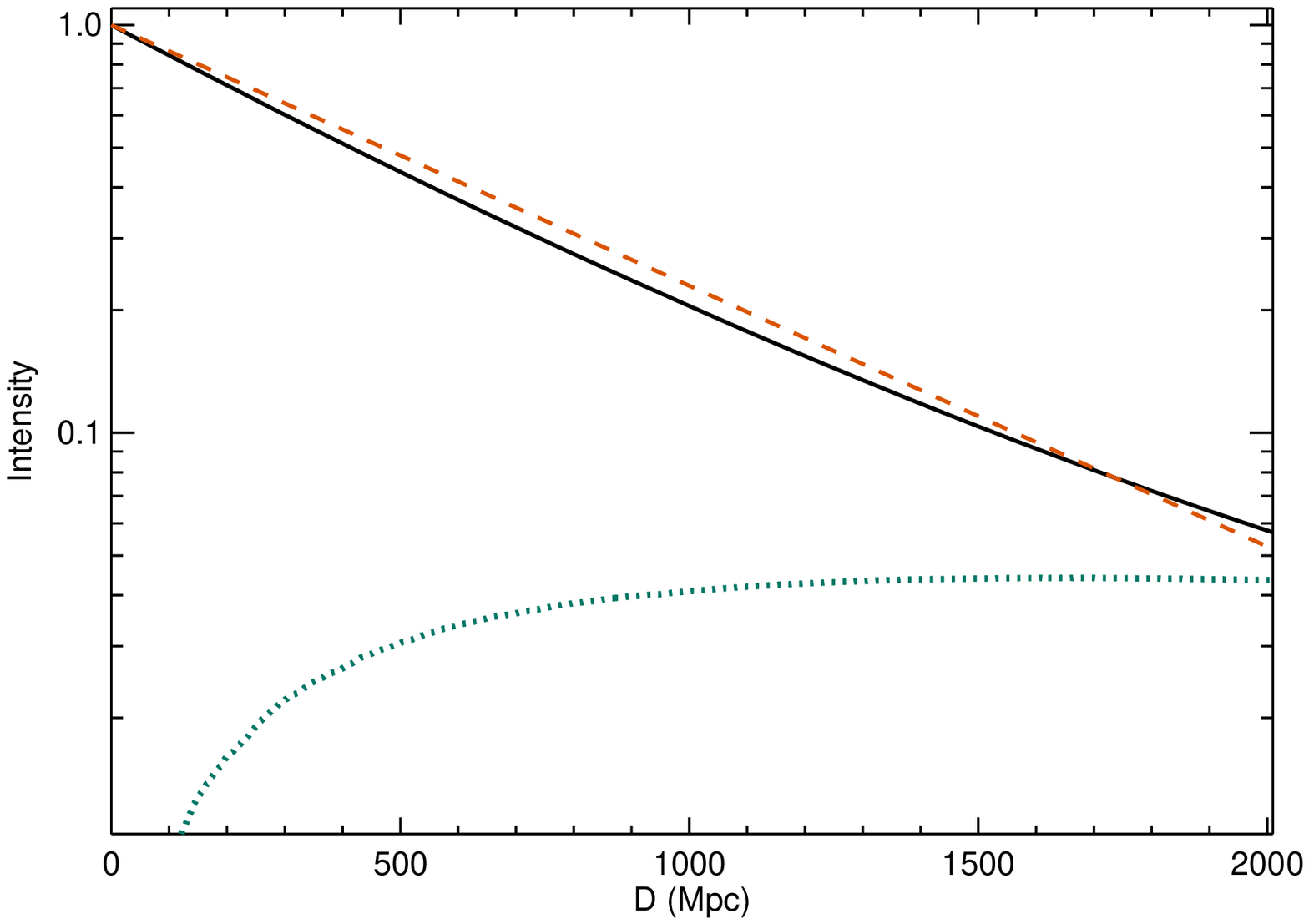}
  \end{minipage}
  \begin{minipage}[b]{0.32\textwidth}
    \centering
    \includegraphics[height=5cm,width=6cm]{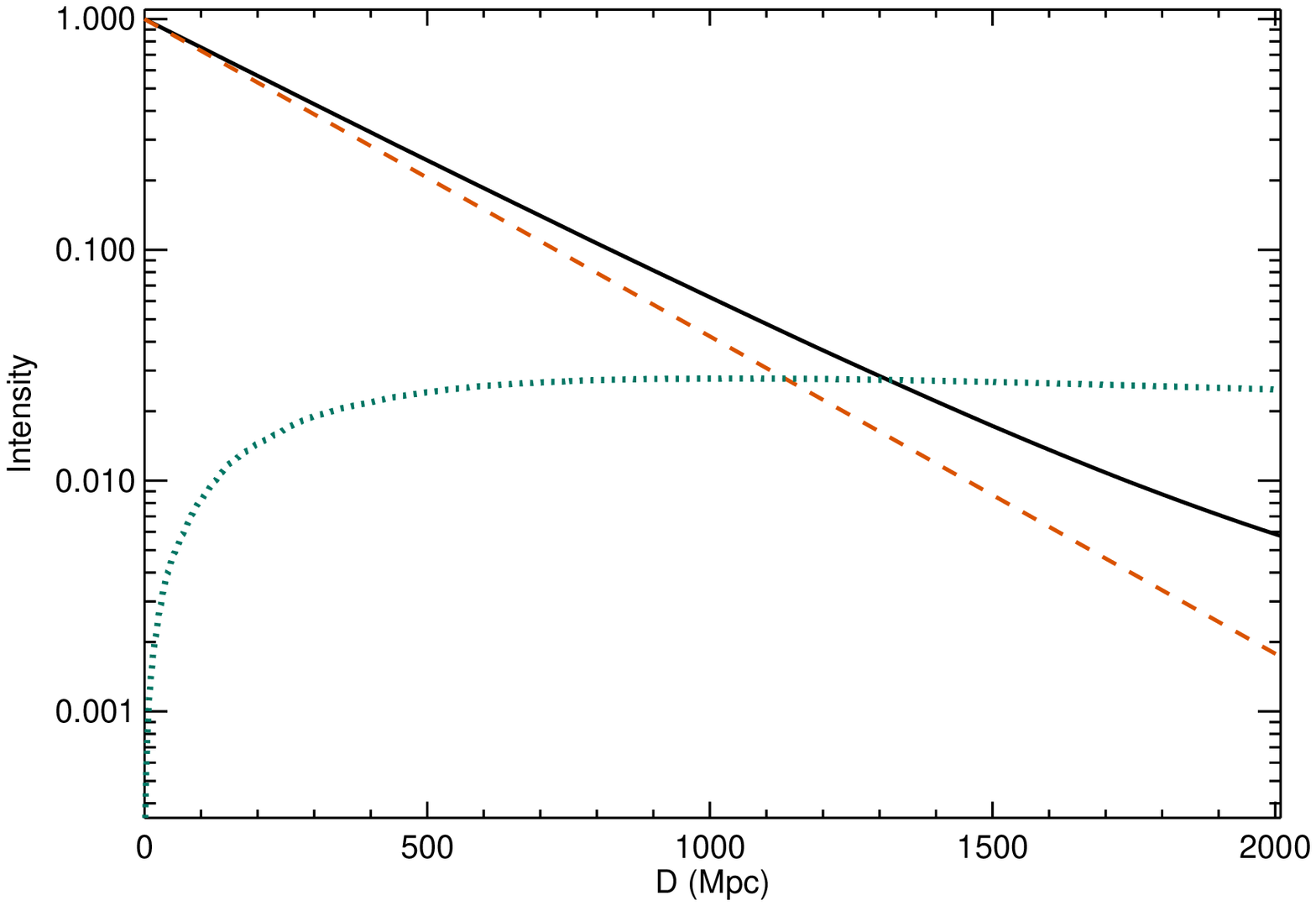}
  \end{minipage}\\
  \begin{minipage}[b]{0.32\textwidth}
    \centering
    \includegraphics[height=5cm,width=6cm]{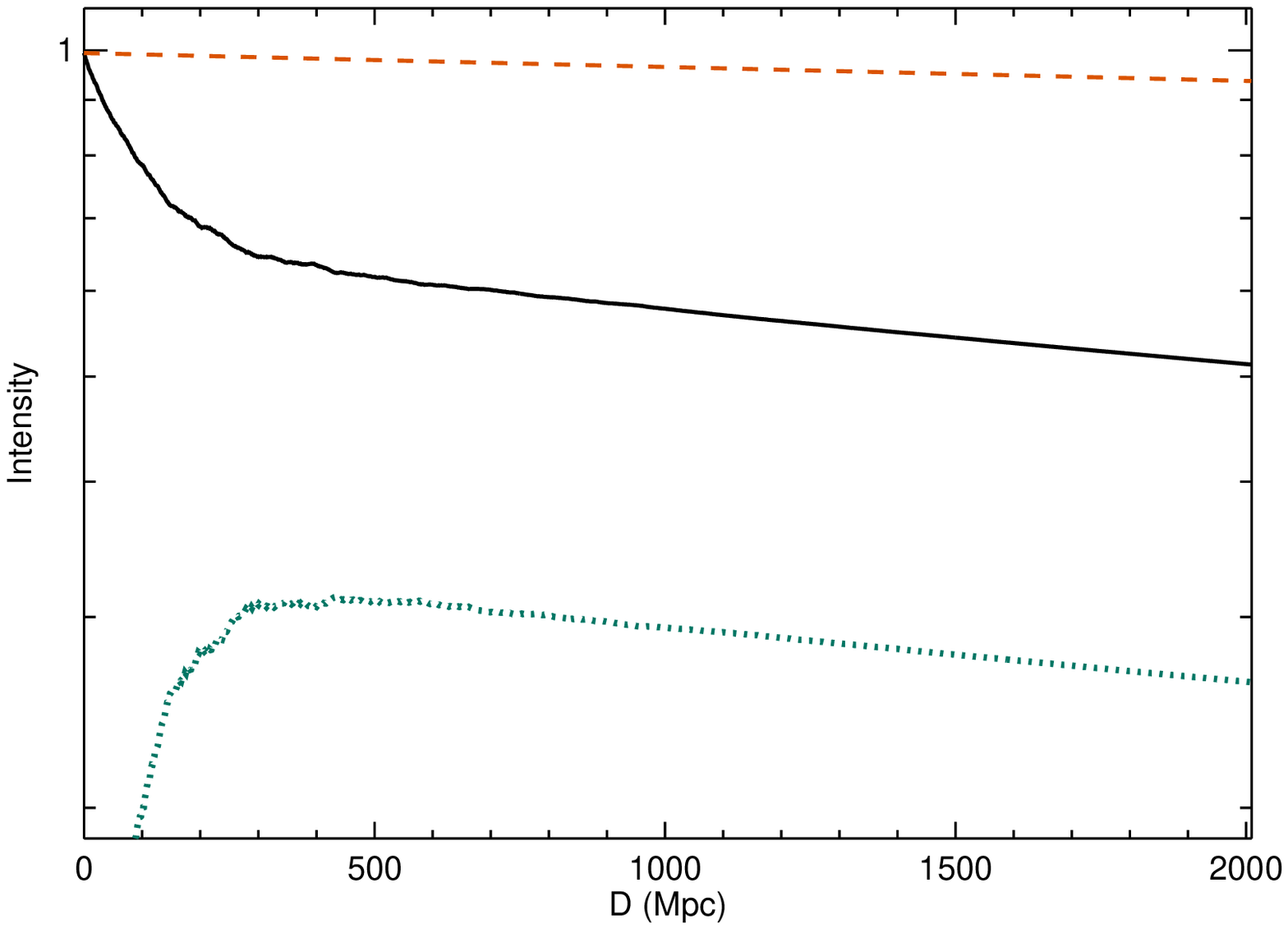}
  \end{minipage}
  \begin{minipage}[b]{0.32\textwidth}
    \centering
    \includegraphics[height=5cm,width=6cm]{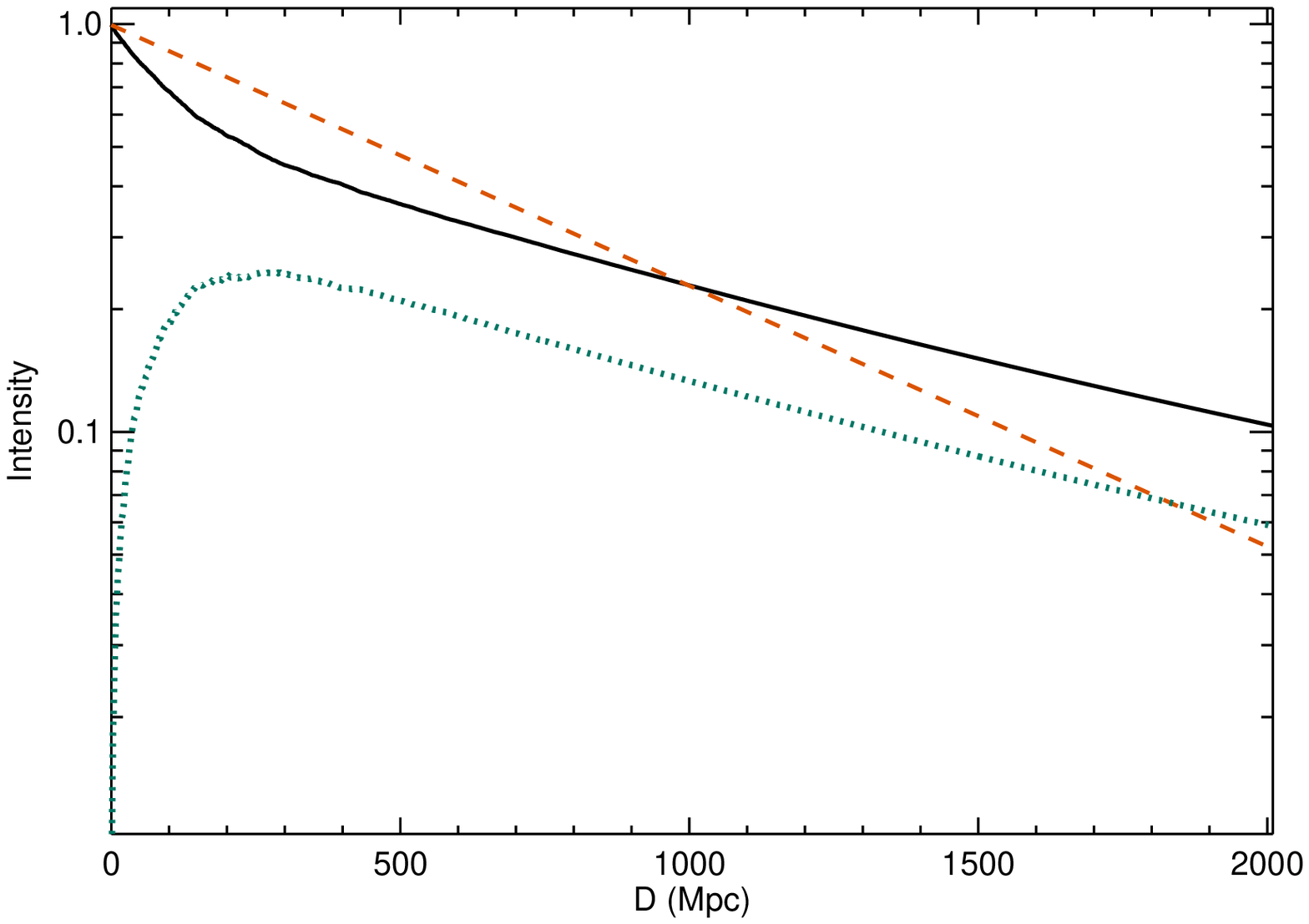}
  \end{minipage}
  \begin{minipage}[b]{0.32\textwidth}
    \centering
    \includegraphics[height=5cm,width=6cm]{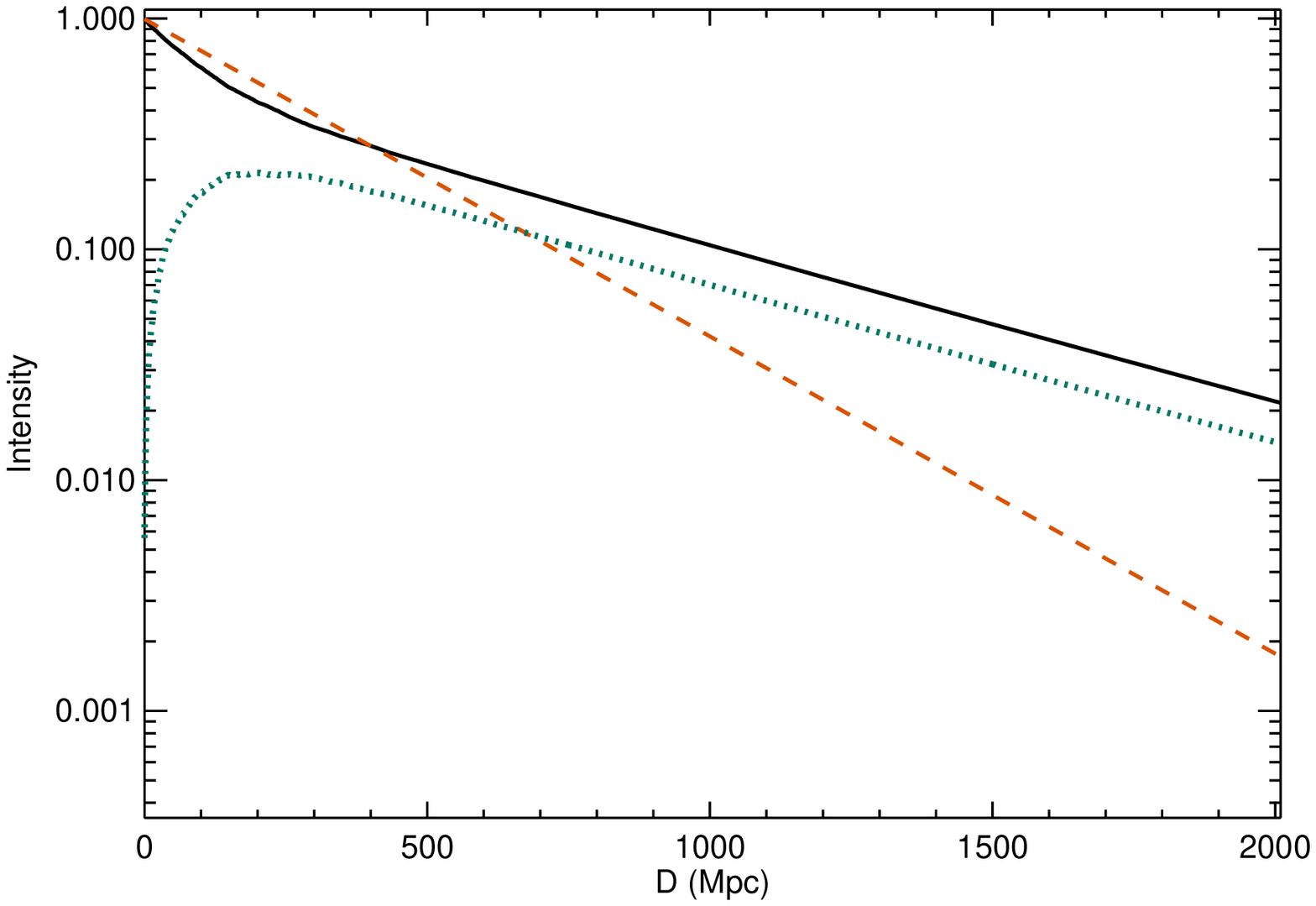}
  \end{minipage}
  \caption{\small{Effect of intergalactic photon/axion mixing on photon and ALP intensities versus distance to the source, computed for our fiducial model, i.e. for 3C~279 and those parameters given in Table~\ref{tab:3c279} but taking B$=1$~nG, and using the Primack EBL model. The black thick solid line represents the total photon intensity, while the blue dotted line is the ALP intensity. The photon intensity as given only by the EBL (i.e. without including photon/axion mixing) is shown as the red dashed line. {\it Top panels}: mixing computed for $M_{11}=4$ GeV and an initial photon energy of 50 GeV (left), 500 GeV (middle) and 2 TeV (right); {\it bottom panels:} $M_{11}=0.7$ GeV and same energies than top panels.}} 
  \label{fig:interg}
\end{figure*}

In our model, we assume that the photon beam propagates over N domains of a given length, the modulus of the magnetic field {\bf B} roughly constant in each of them. We will take, however, randomly chosen orientations, which in practice will be also equivalent to a variation in the strength of the component of the magnetic field involved in the photon/axion mixing. If the photon beam is propagating along the $y$ axis, the oscillation will occur with magnetic fields in the $x$ and $z$ directions since the polarization of the photon can only be along those axis. Therefore, we can describe the beam state by the vector $({\gamma}_x,{\gamma}_z,a)$. The transfer equation will be \cite{csaki03}:

\begin{equation}
\left(\!
\begin{array}{c} 
\gamma_x \\ \gamma_z \\ a
\end{array} 
\!\!\right)
=
{\rm e}^{i E y}
\left[ \, T_0 \, {\rm e}^{\lambda_0 y}
+T_1 \, {\rm e}^{\lambda_1 y} + T_2 \, {\rm e}^{\lambda_2 y} \, 
\right]\!\!
\left(\!
\begin{array}{c}
\gamma_x \\ \gamma_z \\ a                      \label{eq:evolution}
\end{array}
\!\!\right)_{\!\!\!0}   \!
\end{equation}

where:
\begin{eqnarray}
\lambda_0 &\equiv& -\frac{1}{2~{\lambda}_{\gamma}}, \nonumber \\
\lambda_1 &\equiv& -\frac{1}{4{\lambda}_{\gamma}}~\left[ 1 +
\sqrt{1-4~\delta^2} \right] \nonumber \\
\lambda_2 &\equiv& -\frac{1}{4~{\lambda}_{\gamma}}~\left[ 1 -
\sqrt{1-4~\delta^2} \right]
\end{eqnarray}

\begin{widetext}
\begin{eqnarray}
&
T_0 \equiv \left( \begin{array}{ccc}
{\rm sin}^2 \theta & -\, {\rm cos} \theta \, {\rm sin} \theta & 0 \\
-\, {\rm cos} \theta \, {\rm sin} \theta & {\rm cos}^2 \theta & 0 \\
0 & 0 & 0
\end{array} \right) \,\, \qquad
T_1 \equiv \left( \begin{array}{ccc}
\frac{1+\sqrt{1-4\,\delta^2}}{2\,\sqrt{1-4\,\delta^2}}
\, {\rm cos}^2 \theta &
\frac{1+\sqrt{1-4\,\delta^2}}{2\,\sqrt{1-4\,\delta^2}}
\, {\rm cos} \theta \, {\rm sin} \theta &
-\,\frac{\delta}{\sqrt{1-4\,\delta^2}} \,{\rm cos} \theta \\
 \frac{1+\sqrt{1-4\,\delta^2}}{2\,\sqrt{1-4\,\delta^2}}
 \, {\rm cos} \theta \, {\rm sin} \theta &
\frac{1+\sqrt{1-4\,\delta^2}}{2\,\sqrt{1-4\,\delta^2}} \, {\rm sin}^2 \theta &
-\,\frac{\delta}{\sqrt{1-4\,\delta^2}} \,{\rm sin} \theta \\
\frac{\delta}{\sqrt{1-4\,\delta^2}} \,{\rm cos} \theta &
\frac{\delta}{\sqrt{1-4\,\delta^2}} \,{\rm sin} \theta &
-\,\frac{1-\sqrt{1-4\,\delta^2}}{2\,\sqrt{1-4\,\delta^2}}
\end{array} \right) \,\, \nonumber
&
\\
&
T_2 \equiv \left( \begin{array}{ccc}
-\,\frac{1-\sqrt{1-4\,\delta^2}}{2\,\sqrt{1-4\,\delta^2}} \, {\rm cos}^2 \theta &
-\,\frac{1-\sqrt{1-4\,\delta^2}}{2\,\sqrt{1-4\,\delta^2}}
\, {\rm cos} \theta \, {\rm sin} \theta &
\frac{\delta}{\sqrt{1-4\,\delta^2}} \,{\rm cos} \theta \\
-\,\frac{1-\sqrt{1-4\,\delta^2}}{2\,\sqrt{1-4\,\delta^2}}
\, {\rm cos} \theta \, {\rm sin} \theta &
-\,\frac{1-\sqrt{1-4\,\delta^2}}{2\,\sqrt{1-4\,\delta^2}}
\, {\rm sin}^2 \theta &
\frac{\delta}{\sqrt{1-4\,\delta^2}} \,{\rm sin} \theta \\
-\,\frac{\delta}{\sqrt{1-4\,\delta^2}} \,{\rm cos} \theta &
-\,\frac{\delta}{\sqrt{1-4\,\delta^2}} \,{\rm sin} \theta &
\frac{1+\sqrt{1-4\,\delta^2}}{2\,\sqrt{1-4\,\delta^2}}
\end{array} \right) ~
&
\end{eqnarray}
\end{widetext}
$\theta$ being the angle between the $x$-axis and ${\bf B}$ in each single domain. $\delta$ a dimensionless parameter equal to:
\begin{equation}
\label{delta}
\delta \equiv \frac{B \, {\lambda}_{\gamma} }{M} \simeq 0.11 \left( \frac{B}{10^{-9}\, {\rm G}} \right) \left( \frac{10^{11} \, {\rm GeV}}{M} \right)
\left( \frac{{\lambda}_{\gamma}}{{\rm Mpc}} \right)
\end{equation}

\noindent that represents the number of photon/axion oscillations within the mean free path of the photon $\lambda_{\gamma}$. Notice that if there was no EBL, the quanta beam would be equipartitioned between the ALP component and the two photon polarizations after crossing a large number of domains. However, the EBL introduces an energy dependent mean free path $\lambda_{\gamma}$ for the photon.

Amongst all the EBL models that exist in the literature, we chose the Primack \cite{primack05} and Kneiske best-fit \cite{kneiske04} to fix the $\lambda_{\gamma}$ parameter. They represent respectively one of the most transparent and one of the most opaque models for gamma-rays, but still within the limits imposed by the observations (galaxy counts for the lower limit and observations of distant blazars for the upper one). The model proposed by Kneiske et al.~was initially disfavored by some TeV observations of distant AGNs, using the assumption that the intrinsic spectral index needs to be softer than 1.5 (see \cite{Aharonian2006Nature, Mazin2007}). On the other hand, in the literature we also find work where this assumption is strongly criticized, as reported by \cite{Stecker2006HardIndex,Stecker2007HardIndex} and especially in \cite{Krennrich2008}. Therefore, we will consider the Kneiske best-fit EBL model as still valid. In the case of the Primack EBL model, we did not take into account the redshift evolution of the EBL, which effect may be particularly important for sources located at z $>$ 0.3. Although one of the objects we take as the reference, 3C 279, is located at a larger redshift, the difference in the final photon intensity when using the Primack EBL with or without redshift evolution is still small (below 15\% for the relevant energies). This is not true for the Kneiske EBL model, for which the differences might be as high as 70\% for some of the energies under consideration. Therefore, it became necessary to account for such effect in this case.

From each EBL model, we obtain $\lambda_{\gamma}$ as the distance given by the so-called gamma-ray horizon for the energy considered. Additionally, we have to take into account that the energy of each photon will change continuously for a photon traveling towards us from cosmological distances, due to the cosmological redshift. This effect may have a very important role in the calculations of the photon/axion mixing, since e.g. for a source at a distance of 1000 Mpc (i.e. $z\sim0.3$) every photon arrives at Earth with 30\% less energy. We account here for this effect for the first time by computing at each step (distance) the new energy of the photon due to cosmological redshift, and then using this new energy as the input energy needed for the calculation of $\lambda_{\gamma}$. We did not include in the formalism, however, those secondary photons that may arise from the interaction of the primary source photons with the EBL.

To illustrate how the mixing in the IGM works, we show in Figure~\ref{fig:interg} various examples of the evolution of the total photon and ALP intensities as a function of the distance to the source when varying some of the critical parameters, using the Primack EBL model in all cases. We use the parameters listed in Table~\ref{tab:3c279}, that corresponds to our fiducial model, but using an IGMF strength of 1~nG (instead of 0.1~nG), which is still consistent with upper limits. For a photon with an initial energy of 50 GeV (left top panel) and coupling constant $M_{11}=4$, which yields E$_{crit}\sim100$ GeV, there is not significant photon/axion oscillations. Since the role of the EBL is almost negligible at these energies, the total photon intensity remains almost constant traveling to the Earth. For 500 GeV photons (middle top panel), the total photon intensity initially decreases as expected due to the EBL absorption, but also an early extra attenuation due to a photon to ALP conversion is clearly observed. At the same time, the ALP intensity, which was initially equal to zero (we neglect here the mixing inside the source for simplicity), grows rapidly. At larger distances the tendency in the total photon intensity is just the opposite; the intensity increases slightly, since an efficient ALP to photon reconversion (although operative since the very beginning) is taking place and becomes relevant specially at these large distances, where the expected photon intensity is already very low due to the EBL absorption. In the case of photons with higher initial energy (e.g.~2 TeV, right top panel in Fig.~\ref{fig:interg}), the expected attenuation due to the EBL becomes very important even for small distances from the source, which makes more relevant the impact of ALP to photon reconversions on the photon intensity. As a result, the photon/axion mixing implies an enhancement in the photon intensity at almost all distances. The situation changes when using a slightly higher coupling constant, but still within the CAST constraints (see bottom panels in Fig.~\ref{fig:interg}). In this case, both the attenuation and the enhancement in intensity become more pronounced, as expected. For relatively small distances, the photon/axion mixing produces an attenuation in the photon flux, while for relatively large distances, the mixing produces an enhancement in the photon flux. The same argument is essentially valid for any initial photon energy, the results only changing depending on the relative relevance of the EBL in each case, which will modify the distance at which the photon-intensity enhancement starts occurring.

\section{Results}  \label{sec3}

Up to now, previous works have focused only in studying the photon/axion mixing either inside the source or in the IGMFs. Instead, for the first time we carried out a detailed study of the mixing in both regimes under the same consistent framework. We neglect for the moment, however, the mixing that may happen inside the Milky Way due to its galactic magnetic fields. We believe that a concise modeling of this effect is still very dependent on the largely unknown morphology of the {\bf B} field in our Galaxy. In the most idealistic/optimistic case, in which 30\% of the photons convert to ALPs within the source and 10\% of the ALPs convert to photons in the Milky Way, as reported in Ref.~\cite{simet}, this effect would produce an enhancement of the photon flux arriving at Earth of about 3\% of the initial photon flux emitted by the source.

As mentioned in the previous section, in order for the photon/axion oscillation to be observationally noticeable by current instruments, that is $E_{crit} < 1$ TeV, we need ALP masses smaller than $6 \cdot 10^{-10} eV$ for typical values of the IGM. Larger ALP mass values translate into higher $E_{crit}$ (e.g. m$_a$ = 10$^{-6}$ eV would yield $E_{crit}$ $\sim$ PeV in the IGM, when using B $\sim$0.1~nG) making the effect of this oscillation undetectable by the current gamma-ray instruments (Fermi and IACTs). In scenarios with heavy ALPs, then the only effect detectable would be an attenuation caused by photon/axion oscillation in the source, which would be of about 30\% in the most optimistic case \cite{hooper}. It is worth mentioning that the potential photon/axion oscillation in the Milky Way could produce measurable effects only for ALP masses smaller than $10^{-8}$ eV, as mentioned in Ref.~\cite{simet}.

We use an ALP mass of $10^{-10}$ eV in our fiducial model (Table~\ref{tab:3c279}), which implies $E_{crit} \sim$ 30 GeV in the IGM (for B$\sim$0.1~nG) and $E_{crit} \sim$ 1~eV within the source and its vicinity (B$\sim$1~G). Consequently, both effects need to be taken into account.

\subsection{Photon/axion oscillation in our framework} \label{sec:subsec3a}

In this section we show the results obtained when taking into account the mixing inside the source and in the IGMF simultaneously. Since we expect the intergalactic mixing to be more important for larger distances, due to the more prominent role of the EBL, we chose two distant astrophysical sources (as our benchmark AGNs) that are relatively well characterized at gamma-ray energies; namely the radio quasar 3C~279 (z=0.536), most distant detected gamma-ray source at the VHE range, and the BL Lac PKS~2155-304 at z=0.117. In order to compute the photon/axion oscillation in the source we used the parameters reported in Ref.~\cite{hartmann} for 3C~279 and Ref.~\cite{pks2155} for PKS~2155-304. As for the size of the region with ${\bf B}$ field (where photons can convert to ALPs) we chose a region 10 times larger than the radius of the gamma-ray emitting blob given in the above mentioned references, the reason being that the blob radius represents only a lower limit for the region where ${\bf B}$ is confined. We note that this parameter, as well as the number of domains where the ${\bf B}$ is coherent, play an important role in the photon attenuations due to the photon/axion mixing in the source.  Table \ref{tab:atenuaciones} shows the different attenuations that are obtained when varying the size of the region where ${\bf B}$ is confined (what we called ``B region'') and the lengths of coherent ${\bf B}$ domains inside that region. One can see that, once the number of domains is fixed, the photon attenuation increases when increasing the size of the ``B region''. On the other hand, when fixing the size of the ``B region'' and scanning the size of the domains we find that, as we increase the number of domains, the attenuation increases until the size of the domain is ``too small''.  At this point, the probability of photon/axion conversion is almost zero for the single domains, which reduces the overall photon/axion conversion.

\begin{table}[!h]
\begin{center}
  \caption{\label{tab:atenuaciones} \small{Maximum attenuations due to photon/axion oscillations in the source obtained for different sizes of the region where the magnetic field is confined (``B region'') and different lengths for the coherent domains. Only length domains smaller than the size of the ${\bf B}$ region are possible. 
The ${\bf B}$ field strength used is 1.5~G (see Table \ref{tab:3c279}). The photon flux intensity without ALPs was normalized to 1. In bold face, is the attenuation given by our fiducial model.}} 
	\vspace{0.2cm}
    \begin{tabular}{c|ccccccc}
      \hline
      \hline
      B region (pc)    & \multicolumn{7}{c}{Length domains (pc)}\\
      \cline{2-8}
                       & 3$\times 10^{-4}$ & & 3$\times 10^{-3}$ & & 0.03 & & 0.3 \\
      \hline
      0.3               & 0.84 & & 0.67        & & 0.67  & & 0.75 \\
      0.03              & 0.98 & & {\bf 0.84} & & 0.77  & & - \\
      3$\times 10^{-3}$  & 0.99 & & 0.98        & & -     & & - \\
      \hline
      \hline
    \end{tabular}
\end{center}
\end{table}

We summarize in Table \ref{tab:3c279} the parameters we have considered in order to calculate the total photon/axion conversion in both the source (for the two benchmark AGNs) and in the IGM. As we already mentioned, these values represent our fiducial model.

\begin{table}[!h]
\begin{center}
  \caption{\label{tab:3c279} \small{Parameters used to calculate the total photon/axion conversion in both the source (for the two AGNs considered, 3c279 and PKS~2155-304) and in the IGM. The values related to 3C~279 were obtained from Ref.~\cite{hartmann}, while those ones for PKS~2155-304 were obtained from  Ref.~\cite{pks2155}. As for the IGM, e$_{d,int}$ was obtained from \cite{peebles}, and B$_{int}$ was chosen to be well below the upper limit typically given in the literature (see discussion in the text). This Table represents our fiducial model.}} 
  \vspace{0.2cm}
  \begin{ruledtabular}
    \begin{tabular}{l|l|l|l}
      & Parameter & 3C~279 & PKS~2155-304\\
      \hline
      & B (G) & 1.5 & 0.1\\
      Source & e$_d$ (cm$^{-3}$) & 25 & 160\\
      parameters & {\small L domains (pc)} & 0.003 & 3 $\times$ 10$^{-4}$\\
      & B region (pc) & 0.03 & 0.003\\
      \hline
      & z & 0.536 & 0.117\\
      Intergalactic & e$_{d,int}$ (cm$^{-3}$) & 10$^{-7}$ &  10$^{-7}$\\
      parameters & B$_{int}$ (nG) & 0.1 & 0.1\\
      &  {\small L domains (Mpc)} & 1 & 1\\
      \hline
      ALP & M (GeV) & 1.14 $\times$ 10$^{10}$ & 1.14 $\times$ 10$^{10}$\\
      parameters & ALP mass (eV) & 10$^{-10}$ & 10$^{-10}$\\
    \end{tabular}
  \end{ruledtabular}
\end{center}
\end{table}

\begin{figure}[!h]
\centering
  \begin{minipage}[b]{0.49\textwidth}
    \centering
    \includegraphics[height=6.5cm,width=8cm]{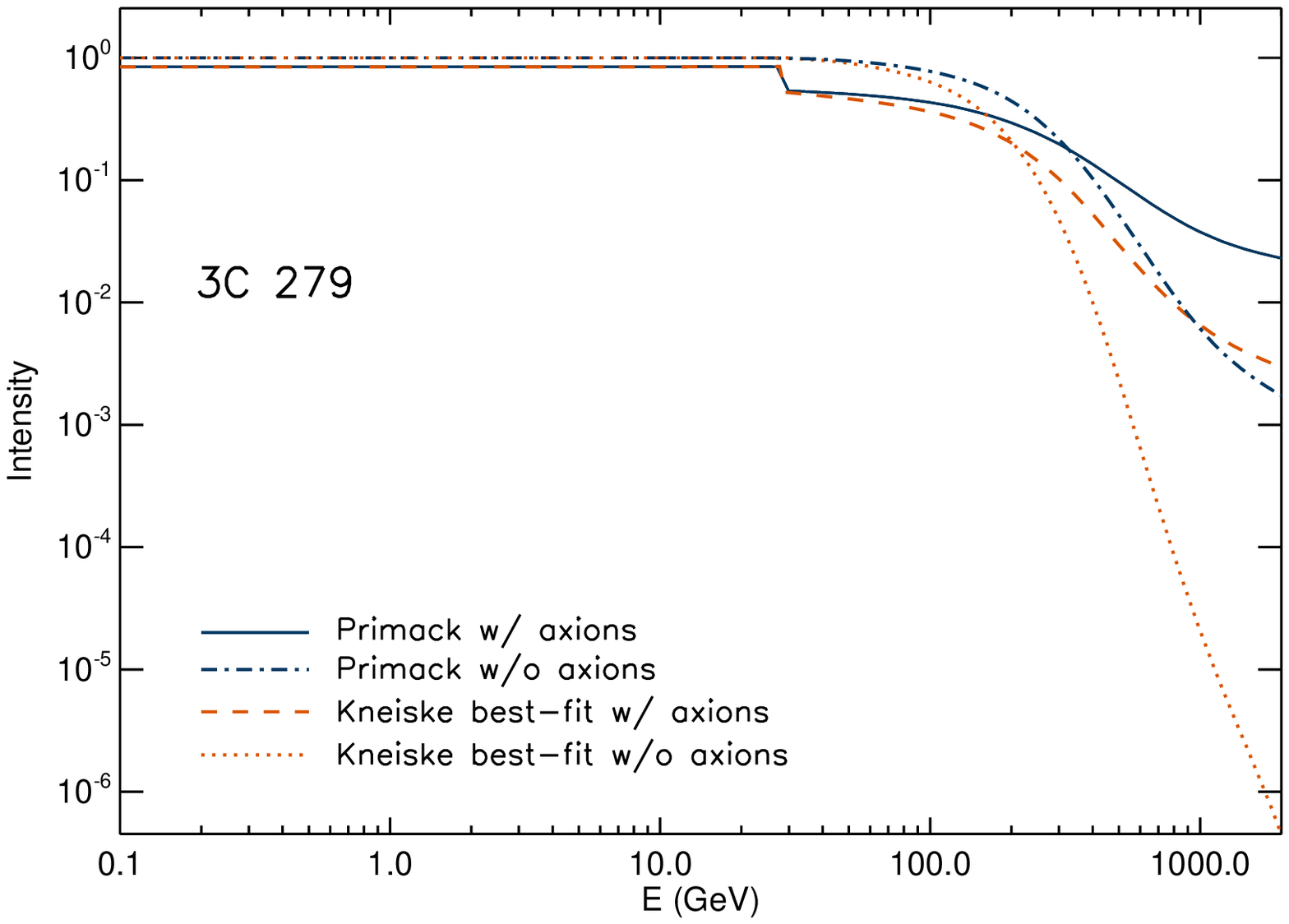}
  \end{minipage}\\
  \begin{minipage}[!h]{0.49\textwidth}
    \centering
    \includegraphics[height=6.5cm,width=8cm]{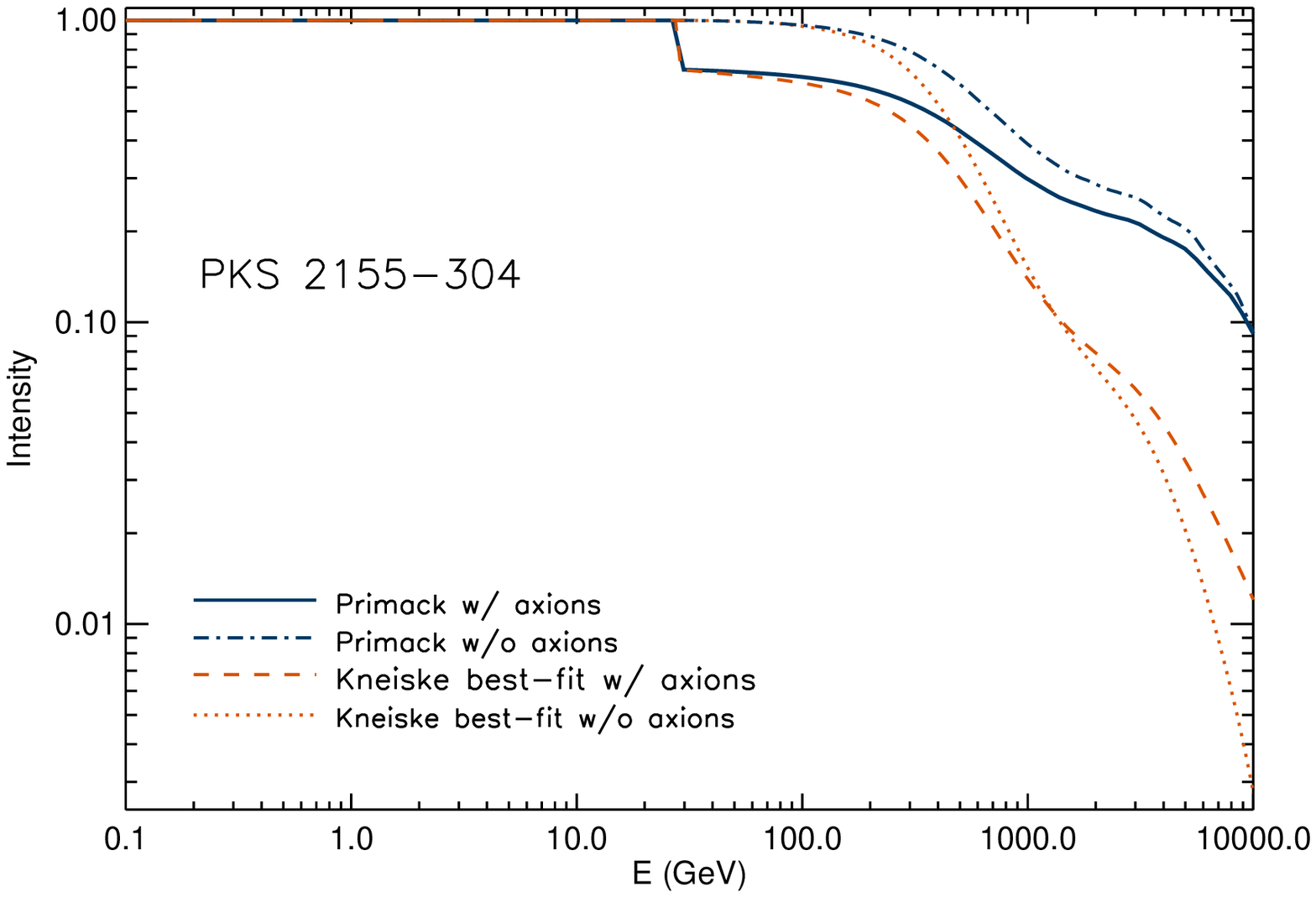}
  \end{minipage}
  \caption{\small{Effect of photon/axion conversions both inside the source and in the IGM on the total photon flux coming from 3C~279 (z=0.536) and PKS~2155-304 (z=0.117) for two EBL models: Kneiske best-fit (dashed line) and Primack (solid line). The expected photon flux without including ALPs is also shown for comparison (dotted line for Kneiske best-fit and dot-dashed line for Primack).}}
  \label{fig:intensity_fiducial}
\end{figure}

The effect of existence of ALPs on the total photon flux coming from 3C~279 and from PKS~2155-304 (using the fiducial model presented in Table \ref{tab:3c279}) can be seen in Figure \ref{fig:intensity_fiducial}. We carried out the calculations for the two EBL models cited above: Kneiske best-fit and Primack. 
The inferred critical energies for the mixing in the source are $E_{crit}=4.6$ eV for 3C~279 and $E_{crit}=69$ eV for PKS 2155-304, while for the mixing in the IGM we obtain  $E_{crit}=28.5$ GeV. The photon attenuation due to photon/axion mixing inside the source is 16\% for 3C~279 and 1\% for PKS 2155-304, as can be seen above their respective critical energies in Figure~\ref{fig:intensity_fiducial}. On the other hand, the photon attenuation due to photon/axion oscillation in the IGM is 30\% for the distance of both sources, and it occurs at the same critical energy. The role of the EBL is negligible at this low energy (i.e. below $\sim$100 GeV), which means that the intensity curves for the two EBL models agree to this energy. 

The situation changes above $\sim$100 GeV, where the photon attenuation due to the EBL is noticeable. At this point, the results depend substantially on the source distance and the EBL model used. A stronger photon attenuation is obtained for the Kneiske best-fit model against the Primack EBL model, as expected. Because the strong photon attenuation due to the EBL, the ALPs that later convert to photons imply a further enhancement of the expected photon flux. Therefore, as one can notice from Figure \ref{fig:intensity_fiducial}, the existence of ALPs translates into a relatively small ($\sim$30\%) intensity attenuation at low energies and a large intensity enhancement (several orders of magnitude, depending on the energy range, distance of the source and chosen EBL model) at high energies.

\begin{figure}[!h]
\centering
  \begin{minipage}[b]{0.49\textwidth}
    \centering
    \includegraphics[height=6.5cm,width=8cm]{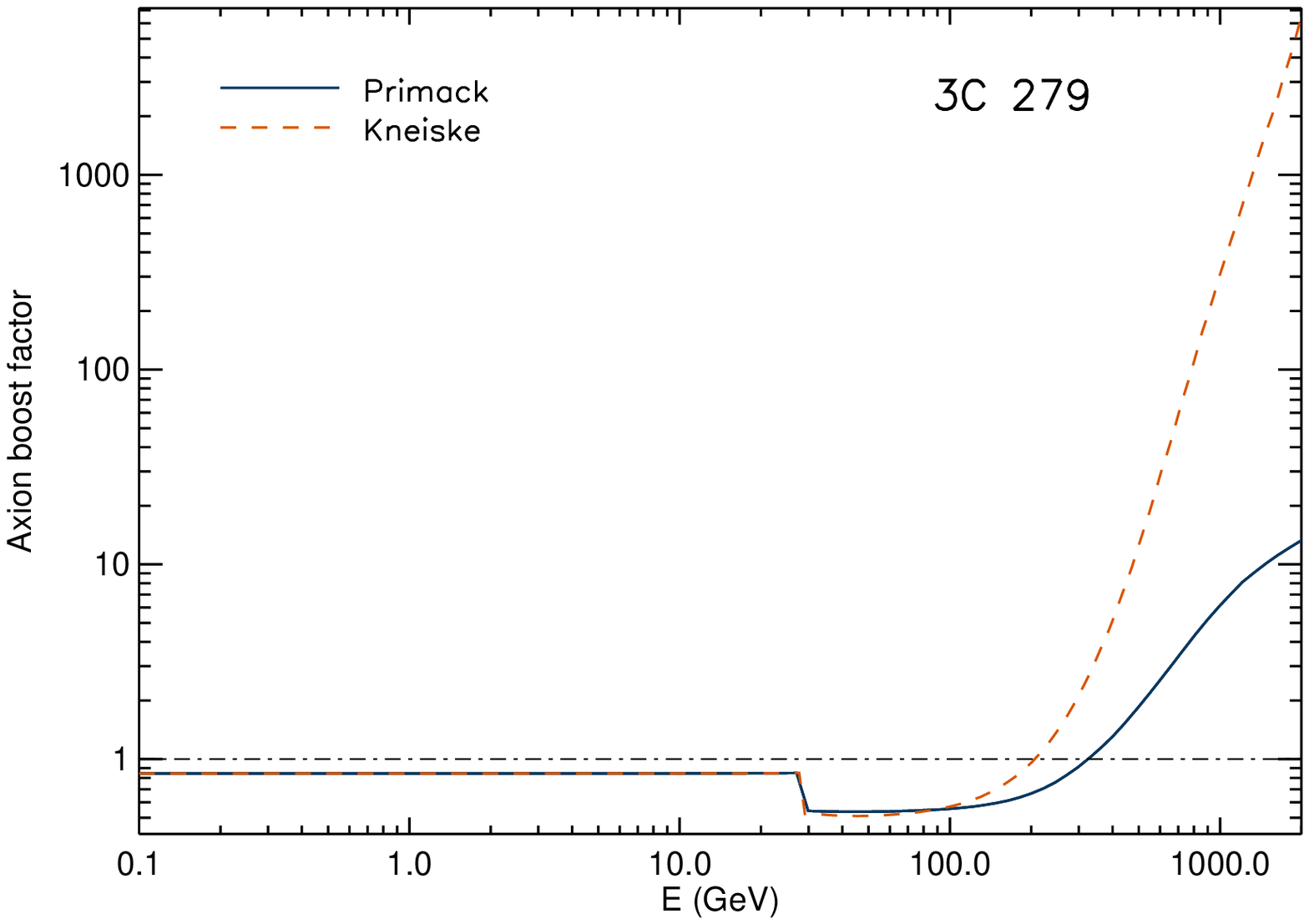}
  \end{minipage}\\
  \begin{minipage}[!h]{0.49\textwidth}
    \centering
    \includegraphics[height=6.5cm,width=8cm]{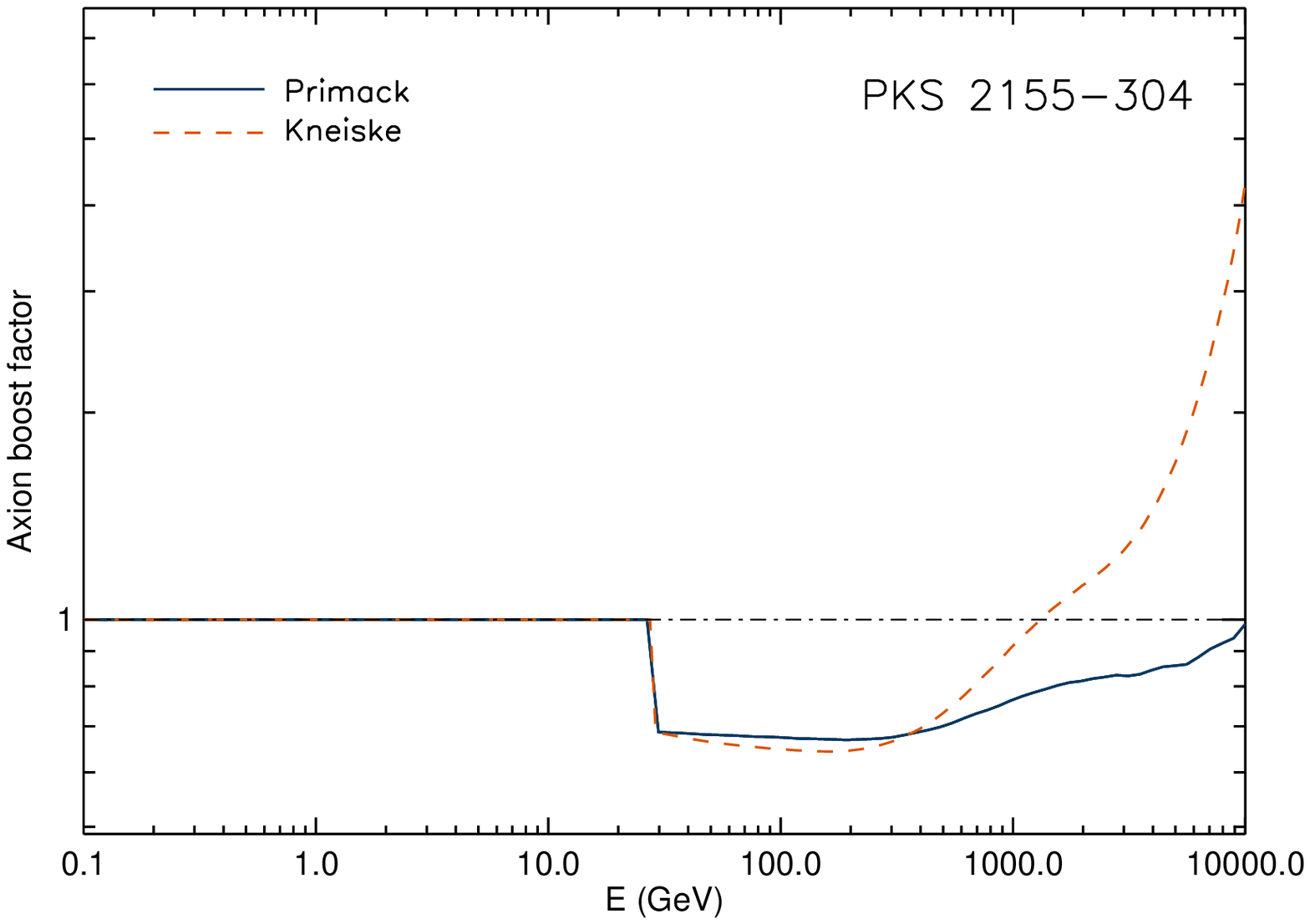}
  \end{minipage}
  \caption{\small{Boost in intensity due to ALPs for the Kneiske best-fit (dashed line) and Primack (solid line) EBL models, computed using the fiducial model presented in Table~\ref{tab:3c279} for 3C~279 (z$=$0.536) and PKS~2155-304 (z$=$0.117).}} 
  \label{fig:boosts_fiducial}
\end{figure}

In order to quantitatively study the effect of ALPs on the total photon intensity, we plot in Figure~\ref{fig:boosts_fiducial} the difference between the predicted arriving photon intensity without including ALPs and that one obtained when including the photon/axion oscillations (called here the {\it axion boost factor}). Again, this was done for our fiducial model (Table~\ref{tab:3c279}) and for the two EBL models described above. The plots show differences in the axion boost factors obtained for 3C~279 and PKS~2155-304 due mostly to the redshift difference. 

In the case of 3C~279, the axion boost is an attenuation of about 16\% below the critical energy (due to mixing inside the source). Above this critical energy and below 200-300 GeV, where the EBL attenuation is still small, there is an extra attenuation of about 30\% (mixing in the IGMF). Above 200-300 GeV the axion boost reaches very high values: at 1 TeV, a factor of $\sim$7 for the Primack EBL model and $\sim$340 for the Kneiske best-fit model. As already discussed, the more attenuating the EBL model considered, the more relevant the effect of photon/axion oscillations in the IGMF, since any ALP to photon reconversion might substantially enhance the intensity arriving at Earth. We note that the axion boost factor may vary when changing the parameters we used to model the source (as shown in Table~\ref{tab:atenuaciones}) and the IGM (see next section). The results we find in this work are in disagreement with those reported by De Angelis et al.~\cite{deangelis}. We {\it always} find that the photon intensity below 200-300 GeV decreases when including the oscillation to ALPs regardless of the ALP and/or IGM parameters, while De Angelis et al. find that the photon intensity increases for a large range of the phase space they tried (see their Fig.~1 in Ref.~\cite{deangelis}). At those low energies the photon attenuation due to pair conversion in the EBL is relatively low (see Fig.~\ref{fig:intensity_fiducial}) and thus the few ALPs that convert to gamma photons do not imply any substantial relative increase in the photon intensity. On the other hand, 1/3 of the photons oscillate to ALPs, which causes a substantial decrement in the amount of gamma photons with respect to those we would have in the absence of ALPs. Therefore, we think it is very difficult to get a photon enhancement at energies $\sim$100 GeV. On the other hand, the axion boost factors we find at high energies ($>$300 GeV) are substantially lower than those obtained in \cite{deangelis}. As an example, in the case of a Kneiske best-fit EBL model with B=1~nG, we find a boost $\sim$4 at 500 GeV, whereas De Angelis et al. obtain $\sim$20 for the same photon energy and the same redshift (note that, in order to carry out a one-to-one comparison with that work, we also used $M_{11}=4$, as they do). One of the reasons for the discrepancy in the axion boost factors is the used EBL model. We noted that the EBL model shown in Fig.~1 of Ref.~\cite{deangelis} is substantially more attenuating than the one from Kneiske best-fit EBL model, which is the one we are using. Consequently, the axion boost factors reported in Ref.~\cite{deangelis} are larger than the ones they would have obtained if they had used the Kneiske best-fit EBL model. Besides that, it is not clear to us whether the change in photon energy due to cosmological redshift (see Section \ref{sec:intergalmix}) was taken into account in \cite{deangelis}; this is not explicitely mentioned in their work.

In the case of PKS~2155-304, the situation is different from that of 3C~279 due to the very low  photon-attenuation at the source and, mostly, due to the smaller source distance. The low redshift location decreases the impact of the EBL absorption and thus the effect of the relative photon-flux enhancement due to photon/axion oscillation. When using B=0.1~nG, the axion boost factor is larger than 1 only for the Kneiske best-fit model and only above 1.3 TeV. In the case of Primack EBL, the axion boost factor is always smaller than 1, thus implying no photon-flux enhancement. Note however that the 30\% drop in the photon intensity occurs at the same energy as that of 3C~279. This drop in the photon intensity should occur at the same energy for all sources located at relatively medium redshifts (0.1$<$z$<$0.3). For very nearby sources (z $<$0.05), the energy drop should still be the same since it only depends on the ALPs properties and the strength of the IGMF. However, the magnitude of the drop will decrease. This is thus a very distinctive and easily testable prediction of this mechanism. We will discuss this issue again in Section \ref{sec4}.

We want to stress another interesting feature that the photon/axion oscillation in the IGMF produces in the source spectra at VHE ($>$100 GeV) of distant sources (z$>$0.1). As one can notice from Fig.~\ref{fig:boosts_fiducial}, the axion boost factor starts increasing at few hundred GeV (when the EBL becomes important), and consequently it will make the source spectra to look harder than they are actually. This happens for both AGNs, yet at slightly different energies: 100 GeV for 3C~279 and 300 GeV for PKS~2155-304. As shown in Figure.~\ref{fig:boosts_fiducial}, the hardening of the VHE spectra occurs for both (very different) EBL models that we used, and hence a very robust prediction of this mechanism being at work. Such a hardening of the spectra was already predicted in Ref.~\cite{simet} for several AGNs located at redshifts 0.1-0.2. While in our work the effect is due mostly to the photon/axion oscillation in the IGMF, in \cite{simet} the effect is due to photon/axion oscillation within the source (up to 30\% attenuation of the photon flux) and the one that occurs in the galactic magnetic fields of the Milky Way (up to 10\% conversion probability). It is worth mentioning here, however, that when using the parameters (essentially ${\bf B}$ strength and size of the ``B region'') for the modeling of the gamma-ray emission from AGN sources, we find that the attenuation in the source due to photon/axion conversion is relatively low; 16\% for our model of 3C~279 and 1\% for that of PKS~2155-304. These low photon-flux attenuation (equivalent to ALP-enhancement) would decrease significantly the effect of the mechanism proposed in \cite{simet}.

Finally, it is worth mentioning that we checked that our results are robust against the randomness of the ${\bf B}$ field. We ran 100 different realizations of the same physical scenario, randomly varying the orientation of ${\bf B}$ in each coherent domain and each realization. We did so for the four cases studied along this work, i.e. 3C~279 and PKS 2155-304, Primack and Kneiske best-fit EBL models. Furthermore, we repeated the same exercise using 0.1, 1 and 10 Mpc as the length of the coherent domains in order to explore the dependence of our results on this parameter. In all cases, we chose our fiducial value of 0.1 nG for the ${\bf B}$ field strength. We found that the maximum differences are typically well below 10\%, implying that the results obtained are not sensitive to the randomness of the B field. We increased the number of realizations to 1000 for some cases and found no differences with respect to the results obtained with 100 realizations.

A larger effect on the computed axion-boost factors occurs when changing the size of the domains being used. The computed axion-boost factors are sensitive to choice of the size of the coherent domains to be used. Together with the choice of the EBL model (which is also uncertain), the choice of the domain sizes modifies the results obtained by factors of a few.

\subsection{The impact of changing B}

\begin{figure*}[!ht]
\centering
  \begin{minipage}[b]{0.49\textwidth}
    \centering
    \includegraphics[height=6.5cm,width=8cm]{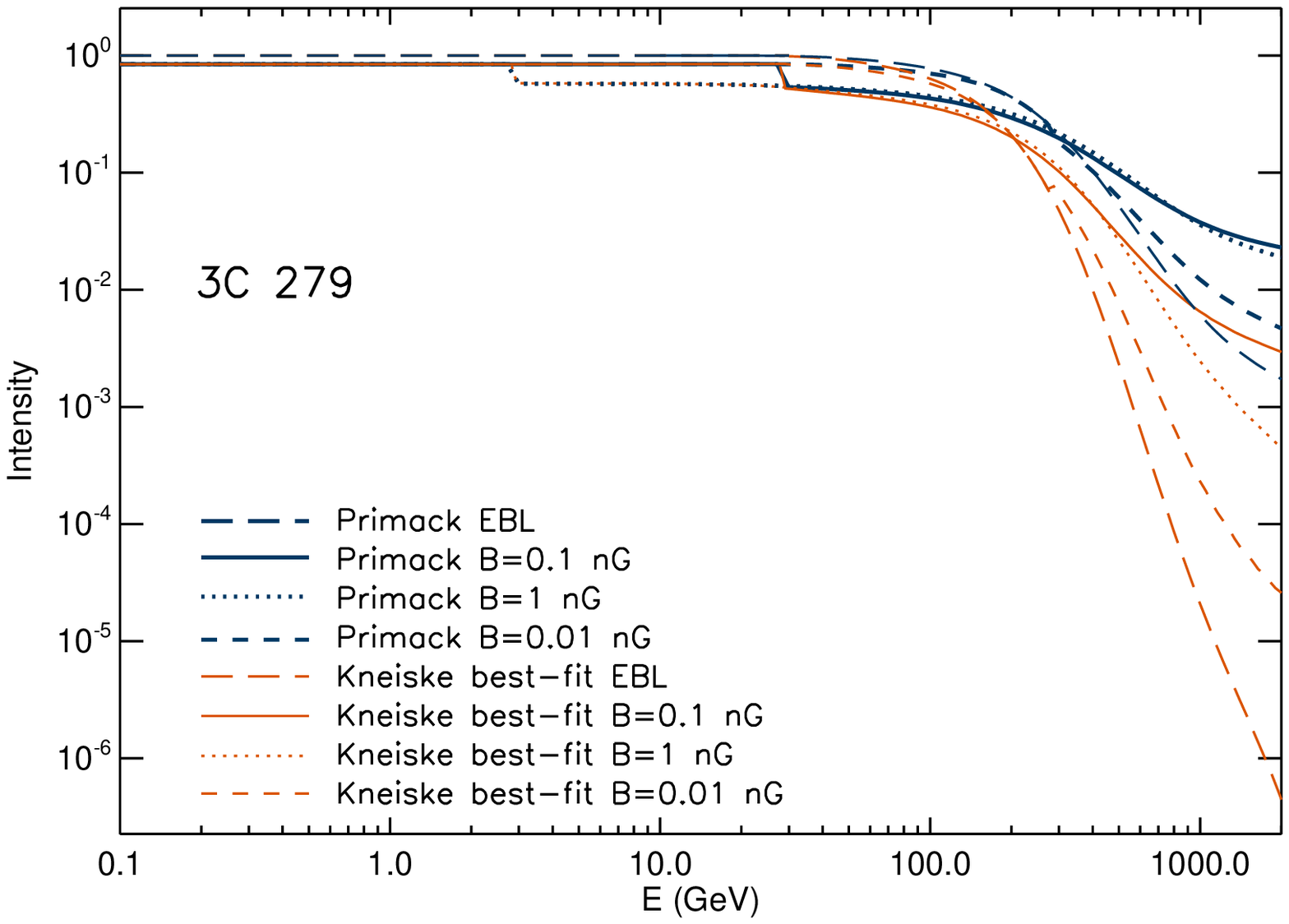}
  \end{minipage}
  \begin{minipage}[b]{0.49\textwidth}
    \centering
    \includegraphics[height=6.5cm,width=8cm]{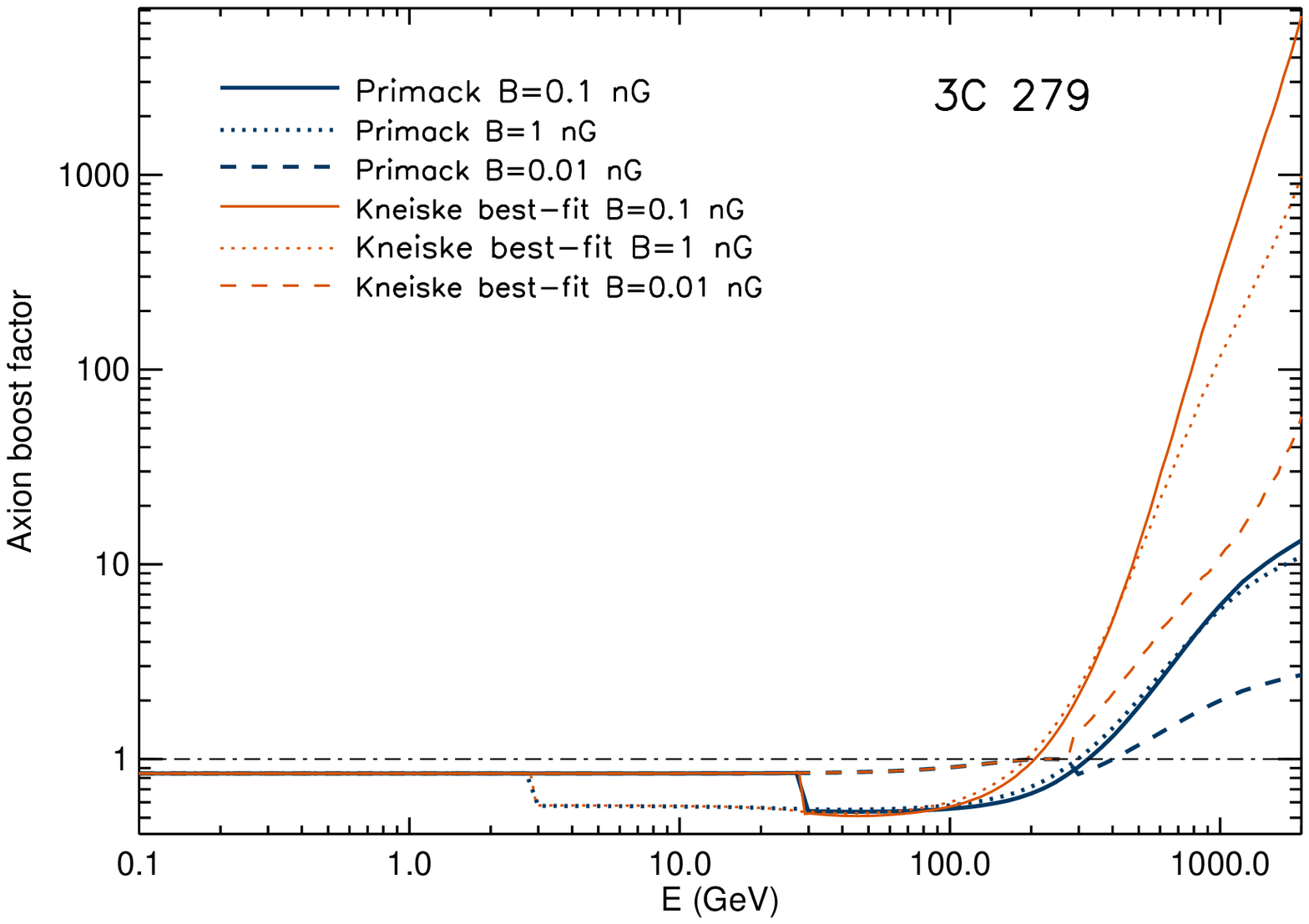}
  \end{minipage}\\
  \begin{minipage}[b]{0.49\textwidth}
    \centering
    \includegraphics[height=6.5cm,width=8cm]{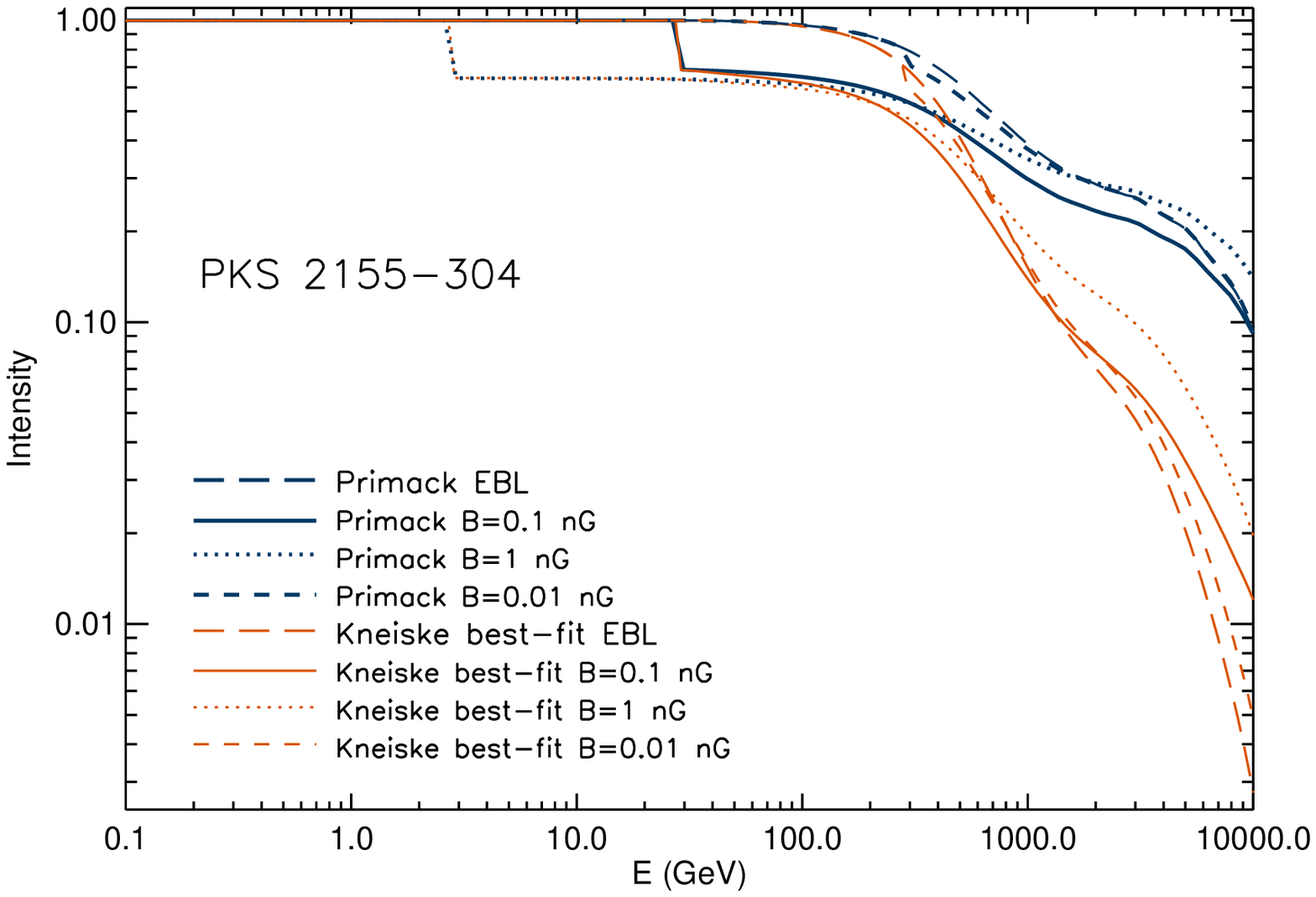}
  \end{minipage}
  \begin{minipage}[b]{0.49\textwidth}
    \centering
    \includegraphics[height=6.5cm,width=8cm]{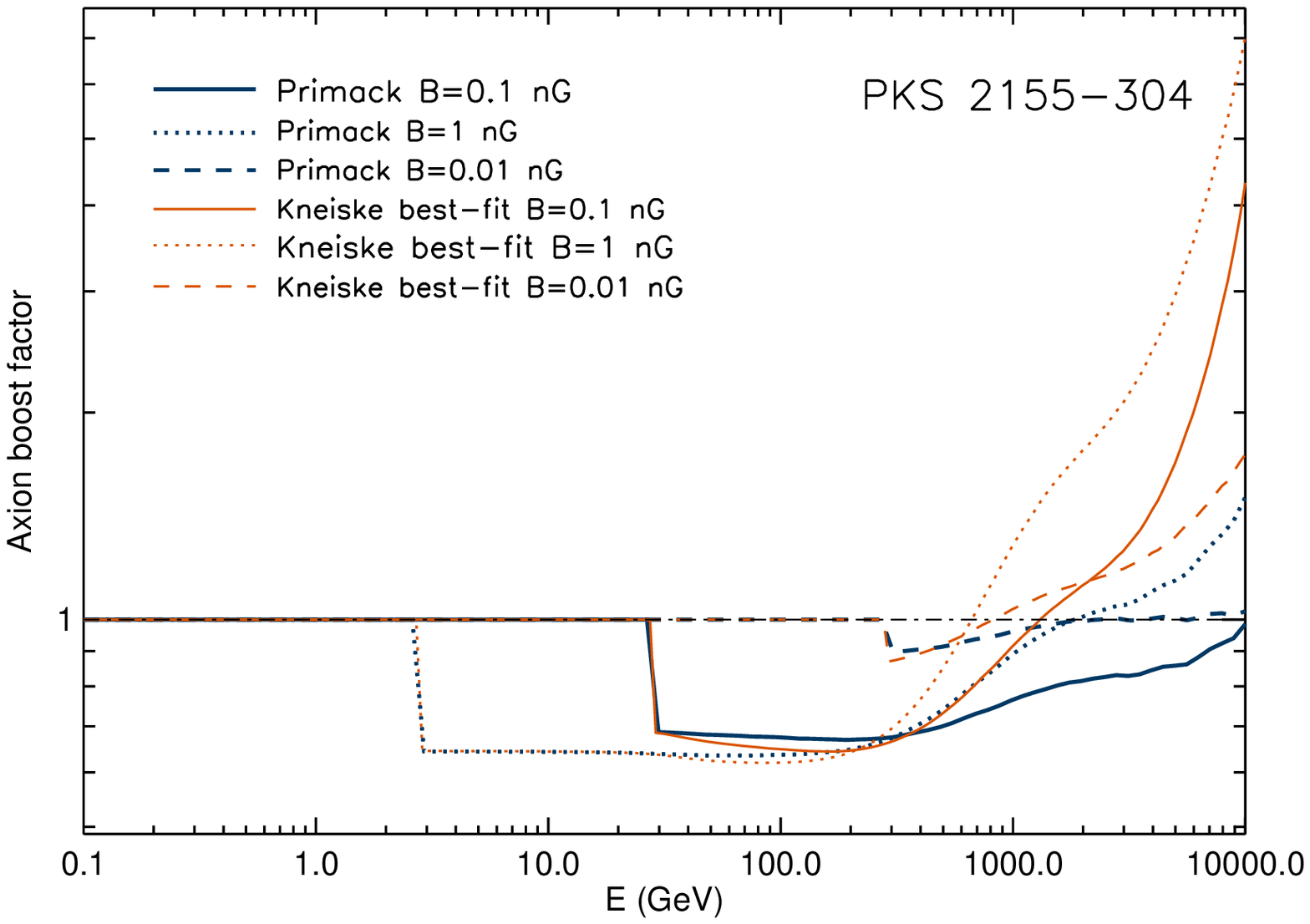}
  \end{minipage}
  \caption{\small{Same as in Figures \ref{fig:intensity_fiducial} and \ref{fig:boosts_fiducial} but for different values of IGMF. Upper panels: 3C~279 using those parameters listed in Table~\ref{tab:3c279}, only changing ${\bf B}$. Lower panels: Same exercise for PKS~2155-304, using the corresponding parameters that can be found in the same Table~\ref{tab:3c279}.}} 
  \label{fig:diffB}
\end{figure*}

A very interesting result has been found when varying the modulus of the intergalactic magnetic field. In Ref.~\cite{simet}, the intergalactic photon/axion mixing was rejected arguing that its effect on the final intensity at Earth would be negligible when using a more realistic value for ${\bf B}$, which should be substantially lower than the value of 1~nG adopted in Ref.~\cite{deangelis}. However, as was shown in the previous section, when using B$=$0.1~nG we find significant effects even for sources located at redshifts as low as z$\sim$0.1. In order to quantify the impact of changing the IGMF strength, we plot in Fig.~\ref{fig:diffB} the result of varying ${\bf B}$ in our fiducial model by one order of magnitude (above and below). We do that for both 3C~279 and PKS~2155-304. 

In the case of 3C~279, we see in the left top panel of Fig.~\ref{fig:diffB} that higher intensities (or equivalently, higher axion boost factors in the right top panel), are obtained when using B$=$0.1~nG instead of taking B$=$1~nG. This seems to contradict the intuitive idea of getting higher intensities for larger magnetic fields, that make the photon/axion mixing more efficient. The reason for this result is the strong attenuation due to the EBL. If the photon/axion mixing is efficient, then many ALPs convert to photons which soon disappear due to the EBL absorption. Consequently, if the source distance is large, we end up having a very small number of photons arriving to the Earth. On the other hand, if the photon/axion mixing is not that strong, then we can keep a higher number of ALPs traveling towards the Earth, which act as a potential reservoir of photons. When decreasing ${\bf B}$ to 0.01~nG, then the axion boost factors are lower than for the other two cases. On the other hand, in the case of PKS~2155-304, we see that the highest axion boost factors are obtained with B=1~nG, because the source is not as distant as 3C~279. If we had considered a source located at a much further distance than 3C~279, then we would have found the highest axion boost factors for B$=$0.01~nG. 

In summary, higher ${\bf B}$ values do not necessarily translate into higher photon flux enhancements. There is always a ${\bf B}$ value that maximizes the axion boost factors; this value is sensitive to the source distance, the considered energy and the EBL adopted model.

\subsection{The impact of using the smallest photon/ALP coupling constant}
The most stringent limits on the ALP-photon coupling constant were derived using the non-detection of gamma-rays (by the Solar Maximum Mission Gamma-Ray Spectrometer) from the supernova (SN) 1987A during the $\sim$10 seconds time window defined by the neutrino burst. This outstanding event allowed several authors in 1996 to set lower limits to the inverse of the coupling constant $M_{11}$ to values larger than 1 \cite{brockway1996} and 3 \cite{griffols1996}. Those limits are only valid for ultralight ALPs. In both works the value $m_{a} < 10^{-9}$ eV is quoted, although this value holds only for some specific situations. Indeed, a more robust value is $m_{a} < 10^{-11}$ eV (see Refs.~\cite{hooper,simet}), i.e. the energy below which the exact value of the ALP mass is irrelevant because the ``plasma frequency'' dominates (see definition of ALP effective mass in Section \ref{sec_source}). Various authors (see Refs.~\cite{csaki03,deangelis}) used $M_{11}=4$ when dealing with $m_a < 10^{10}$ eV. Since the ALP mass in our fiducial model is $10^{-10}$ eV, and hence close to this limit, we decided to repeat the calculations using this value for $M$, which is 35 times larger than the value we used in the previous sections (see Table~\ref{tab:3c279}).  

Before we continue, it is worth pointing out that the limits to the ALP-photon coupling constant given in Refs.~\cite{brockway1996,griffols1996} are subject to large uncertainties that are not fully discussed in those papers. Both the flux of ALPs produced in the SN explosion and the back-conversion of ALPs to gamma photons can vary by large factors, and hence the upper limits computed with those numbers have to be taken with caveats. 

The calculated flux of ALPs produced and released during the SN explosion depends on the knowledge of the size, temperature and density of the proto-neutron star. Those numbers are subject to large uncertainties because we still do not know how stars explode. Even though there is general agreement that the ultimate energy source is gravity, the relative roles of neutrinos, fluid instabilities, rotation and magnetic fields continue to be debated. In particular, back in the 90s it was believed that neutrinos would be able to reheat the outgoing shock-wave and produce the explosion. Nowadays, with far more powerful computer simulations, we know that neutrino-driven explosions are only possible when the star has a small iron core and low density in the surrounding shells, as being found in stars near or below 10 solar masses \cite{Janka2007}. The progenitor of SN1987A was a blue supergiant and hence it is expected to be somewhere between 10-50 solar masses. A possibility to explain those explosions might require the proper inclusion of rotation and magnetic fields (see Refs.~\cite{Burrows2007,Obergaulinger2008,Dimmelmeier2008} and references therein). Both B field and rotation are present in stars as well as in pulsars, which are the products of successful SN explosions; thus it is very natural to consider them in SN explosion models. In particular, the rotation of the proto-neutron star can change substantially the temperature and, specially, the density of the inner core; in \cite{Dimmelmeier2008} it is shown that the density can vary by more than one order of magnitude, which would change by a similar factor the flux of ALPs being produced. Refs.~\cite{brockway1996,griffols1996} did not consider such level of complexity (and uncertainties) in the parameters used to compute the flux of ALPs, mostly because 15 years ago we lacked that knowledge.

On the other hand, the back-conversion of ALPs to photons relies on the structure of the galactic magnetic field which is, again, not well known. Different models predict B fields that could differ substantially and hence they would predict different values for the amount of gamma photons we would obtain for a given flux of ALPs. This is clearly shown in Fig.~1 from Ref.~\cite{simet}, where the probability of ALP-photon conversion is given for various locations of the sky. Therefore, even if we could accurately predict the number of ALPs from SN1987A, the number of photons would be subject to large uncertainties. 

\begin{figure}[!h]
\centering
  \begin{minipage}[b]{0.49\textwidth}
    \centering
    \includegraphics[height=6.5cm,width=8cm]{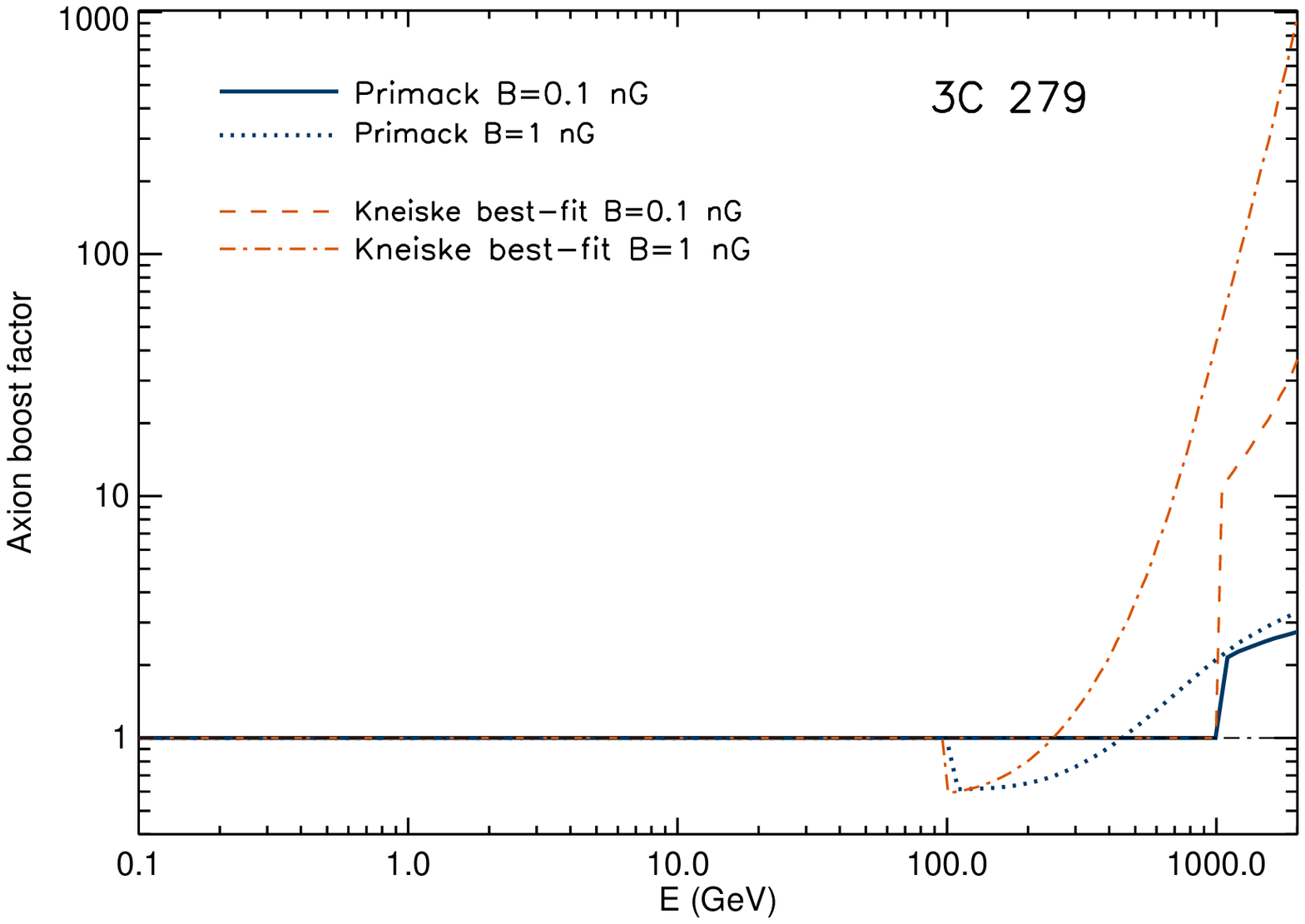}
  \end{minipage}\\
  \begin{minipage}[!h]{0.49\textwidth}
    \centering
    \includegraphics[height=6.5cm,width=8cm]{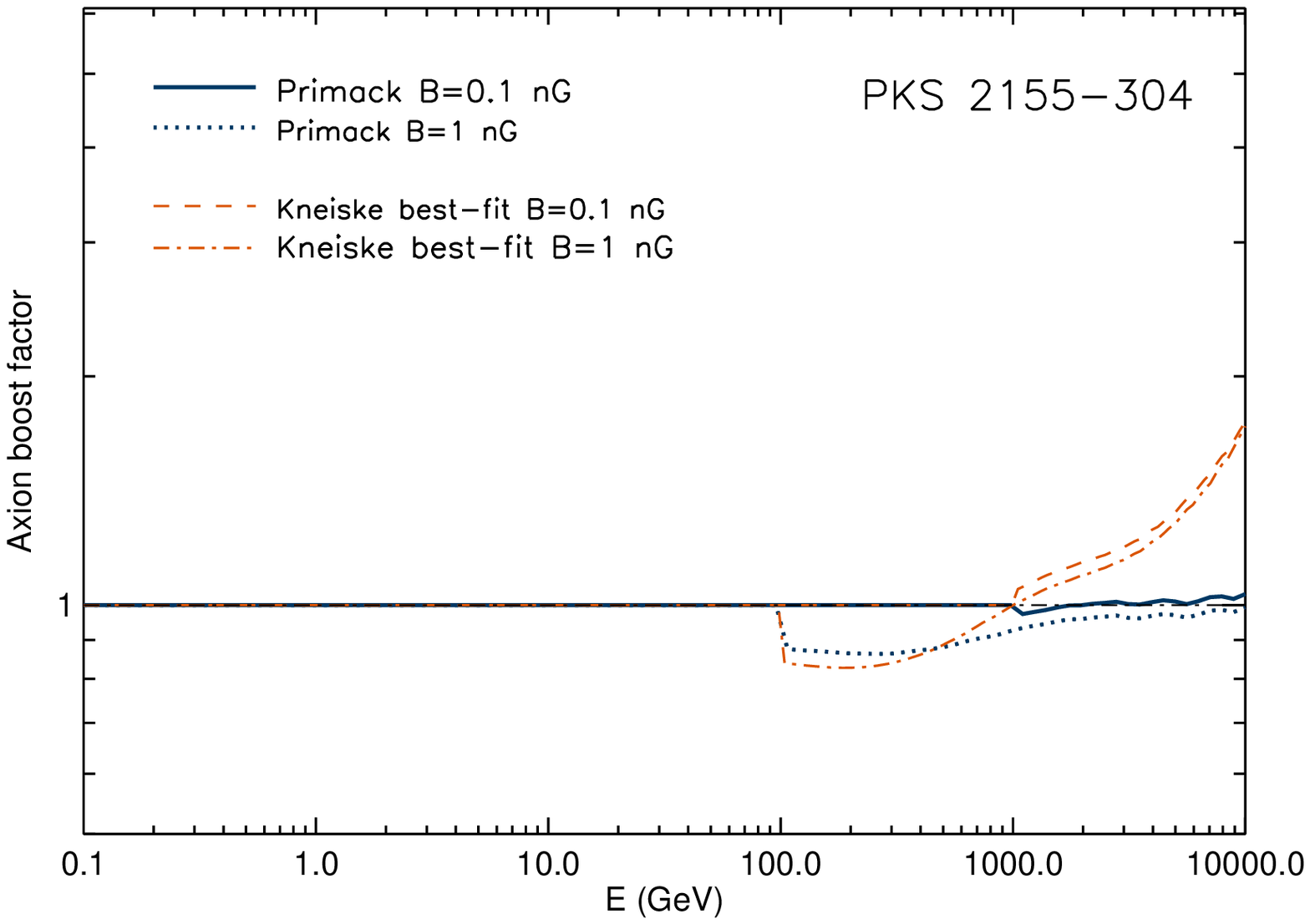}
  \end{minipage}
  \caption{\small{Boost in intensity due to ALPs for the Kneiske best-fit and Primack EBL models, computed using the fiducial model presented in Table~\ref{tab:3c279} for 3C~279 (z$=$0.536) and PKS~2155-304 (z$=$0.117), but with $M_{11}=4$ GeV} and B=0.1~nG (dashed and solid lines for Kneiske best-fit and Primack EBL models respectively) and B=1~nG (dot-dashed and dotted lines).} 
  \label{fig:boosts_fiducial_NewM}
\end{figure}

Therefore, we conclude that the limit in the inverse of the ALP-photon coupling constant given in Refs.~\cite{brockway1996,griffols1996} is subject to large (orders of magnitude) uncertainties, and thus the limit given by the CAST collaboration remains as the most robust one up to date. However, for the sake of comparison with other works, we computed the axion-boost factors when using $M_{11}=4$ eV. This is shown in Figure \ref{fig:boosts_fiducial_NewM} for both 3C~279 and PKS~2155-304 for two values of the B field, 0.1 nG and 1 nG. For this low coupling constant, the effect due to the photon/ALP oscillation in the source is negligible. The effect due to photon/ALP oscillation in the IGMF in not negligible, but substantially lower than the one shown in the previous section. Besides, such effect shows up at larger energies now (see Eq. \ref{eq:ecrit}); 100 GeV and 1000 GeV respectively for 1~nG and 0.1~nG.

\section{Detection prospects for Fermi and IACTs}  \label{sec4}

As mentioned in Section \ref{sec2}, for photon/axion coupling constants close to the current published limits, and for realistic ALP mass values, the energy at which the photon/axion oscillation starts to become important is expected to lie in the gamma-ray range. Consequently, the combination of the Fermi/LAT instrument and the IACTs, which cover 6 decades in energy (from 20 MeV to 20 TeV) is very well suited to study the photon/axion mixing effect. Because of the rapid change in the predicted photon intensity attenuation close to $E_{crit}$, the energy resolution of the instrument is very relevant in order to detect the ALP signatures. The Fermi/LAT instrument has an energy resolution of about 10\%, whereas IACTs, above 150 GeV, have an energy resolution of about 20-25\%. The photon intensity attenuation we found in this work goes from 0 to 30\% due to the mixing in the source, plus essentially 30\% due to the mixing in the IGM. This implies that one needs to be able to determine photon fluxes with a precision better than 10\%. Such level of precision might not be achievable for energies $>$10 GeV with Fermi, or for energies $>$1 TeV with IACTs due to a low photon counting (which depends obviously on the brightness and hardness of the gamma-ray sources). On the other hand, if the source is emitting at $\sim$TeV energies and is located at large distances, the photon/axion oscillation in the IGM could translate into an intensity enhancement of more than one order of magnitude (see Figures \ref{fig:intensity_fiducial}, \ref{fig:boosts_fiducial}, \ref{fig:diffB}), which should be certainly easy to detect with current IACTs. 

Therefore, if we accurately knew the intrinsic spectrum of the sources and/or the density of the EBL, we should be able to observationally detect ALP signatures for a wide range of the parameter space (photon/axion coupling constant and ALP mass). The main problem is that we do neither know accurately the intrinsic spectrum of the sources nor the EBL density. Thus the potential detection of those ALP signatures become quite challenging, but not impossible. In order to study this scenario, we propose the following strategy:

\begin{enumerate}
\item Observe several AGNs located at different redshifts, as well as the same AGN undergoing different flaring states (low/high fluxes), at different energy ranges, from radio to TeV. This is important because the modeling of the gamma-ray emission depends critically on the emission at lower energies (specially infrared, optical, UV and X-rays), and also because we do not know {\it a priori} the energy at which the photon/axion oscillation will start to operate.

\item Try to describe the observational data with ``conventional'' theoretical models for the broad band emission (Synchrotron Self-Compton, External Compton, Proton synchrotron, etc) and the attenuation of the gamma-rays in the EBL (Primack, Kneiske best-fit or other EBL models). The current models (sometimes very simplistic) will definitely require some modifications to fit the observational data.

\item Look for intensity drops in the residuals (``best-model''-data).  We want to stress that the drop in the photon flux due to the attenuation in the IGM only depends on the IGMF and the properties of the ALPs (mass and coupling constant), i.e.~it is independent of the gamma-ray sources. Therefore, a detection of such photon flux drops at the same energy in a numerous of different sources  would be a clear signature for the existence of photon/axion oscillation, because we do not expect that the intrinsic spectrum from different sources (or same source at different flux levels) have a rapid drop of $\sim$30\% in the emission at the same energy. The detection (or no-detection) of this photon intensity drop implies a constraint for the product $m_a^2 \cdot M_{11}$. In this specific search, the Fermi-LAT instrument is expected to play a key role since it will detect thousands of AGN sources located at various redshifts (up to z$\sim$5), and at energies where the photon absorption due to the EBL is not important. 

\item Look for intensity enhancements in the residuals (i.e. ``best-model''-data). This should occur at the highest energies (E$>$300 GeV) and thus only detectable with IACTs. The origin of the potential photon flux ``excess'' might be due to a wrong EBL model and/or wrong model for the source emission, the last being very important because it introduces differences between the different sources (or the same source under different activity levels). In this case what we need is to detect distant (z $>$ 0.2) sources at the highest possible energies ($>$1 TeV). The current EBL models are already very close to the minimum possible photon density limits from galaxy counts; that is, we cannot make them much more transparent. That implies that the detection of TeV photons from a source that is at redshift 0.5 (like 3C~279) could not be explained with conventional physics, regardless of the intrinsic spectrum of the source. This would be a strong hint for the existence of photon/axion oscillation. If the same effect was observed for different sources at different redshifts, we could try to parameterize the effect by varying the ALP parameters (and/or the IGMF strength). If that parameterization could be done successfully, then we would not only have a very strong hint for the existence of ALPs, but also would be able to constrain the available parameter space (coupling constant and ALP mass).
\end{enumerate}

The detection of ALPs is not trivial (and cannot be done with just few sources), but it is certainly possible, as we have shown above. Along these lines, it is worth mentioning that we might be already starting to see hints of the existence of ALPs from the gamma-ray spectra of cosmological sources. Very recent works already pose substantial challenges to the conventional interpretation of the observed source spectra from several distant AGN sources. On the one hand, the VERITAS Collaboration recently claimed a detection of gamma-rays above 0.1 TeV (the highest detected energies are not yet reported) coming from 3C~66A \cite{Swordy2008}, an intermediate-frequency-peaked BL Lac object located at redshift 0.444. This claim coincides with the detection, at GeV energies, of this source in active state by the Fermi-LAT instrument during the same time window \cite{Tosti2008}. In addition, the MAGIC Collaboration reported gamma-rays above 1 TeV coming from a location consistent with the position of 3C~66A \cite{Albert2008_3c66a}. Those observations would confirm the earlier claims from the Crimean Astrophysical Observatory’s GT-48 IACT of several detections from this source above 0.9 TeV \cite{Nesphor1998,Stepanyan2002}. Those detections were not confirmed by the HEGRA and Whipple telescopes, which are more sensitive instruments, but which observed the source at different time windows \cite{Aharonian2000,Horan2004}. As mentioned above, a detection of TeV photons from a source located at z=0.444 would pose serious problems to conventional models of photon propagation over cosmological distances, where the high energy gammas are expected to disappear due to pair electron-positron production in the EBL. On the other hand, the recent published lower limits to the EBL at 3.6 microns \cite{Levenson2008}, which is almost twice larger as the previous ones, enhances even further the attenuation of gamma-rays at TeV energies and thus increases even more the magnitude of the mystery. Furthermore, as reported in \cite{Krennrich2008}, this fact extends the problems to sources located at medium redshifts (z=0.1-0.2) whose intrinsic energy spectra appear to be harder than previously anticipated. Those observations present blazar emission models with the challenge of producing extremely hard intrinsic spectra (differential spectral index in the spectrum smaller than 1.5) in the sub-TeV to multi-TeV regime. As mentioned in the previous section, the photon/axion oscillation in the IGM would naturally explain these two puzzles; the detection of TeV photons from very distant (z$\sim$0.5) AGNs, and the apparent hardening of the spectra for relatively distant (z$>$0.1) AGNs.

However, it is worth mentioning that the above reported puzzles might still be explained with conventional physics, as well as uncertainties in the published numbers. 
The measured redshift of 3C~66A could be wrong \cite{Bramel2005,Finke2008}, or the TeV photons reported by MAGIC could come from a neighboring source (a radio galaxy), 3C~66B, which has never been detected in gamma-rays \cite{Albert2008_3c66a}.  As for the blazars with intrinsic spectra harder than 1.5, there is currently quite some controversy. Some authors claim that spectra harder than 1.5 could be possible (see e.g. \cite{Katar2006,Stecker2007,Aharonian2008,Boettcher2008}), while others state that spectra should be always softer than 1.5 (see for instance \cite{Malkov,Aharonian2006,Boettcher2008,3c279magic}), and use this value to set upper limits to the EBL density at infrared frequencies \cite{Coppi1999,Aharonian2006,Mazin2007,3c279magic}. An argument in favor of the latter is the fact that EGRET never measured spectra harder than 1.5  at energies below 10 GeV, where the EBL does not distort the gamma-ray spectra, for any of the almost 100 detected AGNs \footnote{In this matter, the Fermi-LAT instrument, with a sensitivity one order of magnitude better, and scanning the complete sky every 3 hours, is expected to see thousands of AGNs; which will surely shed some light into this mystery.}.

Finally, we would like to note that the capabilities of detecting the mentioned signatures will increase significantly with the new generation of ground instruments, i.e. MAGIC~II or HESS~II (with lower energy thresholds, expected to operate in 2009), CTA and AGIS (with even lower energy thresholds and higher sensitivity at multi-TeV energies) and HAWC (higher sensitivity at multi-TeV energies, large duty cycles).

\section{Conclusions}  \label{sec5}

If ALPs exist, then we should expect photon to ALP conversions (and vice-versa) in the presence of magnetic fields. This photon/axion mixing will occur in gamma-ray sources as well as in the IGM. We have explored in detail both mixing scenarios together in the same framework. The main conclusions on this work can be summarized as follows:

\begin{itemize}
\item If photons oscillate into ALPs in the IGM, then photon/axion mixing in the source is also at work for lower photon energies. In this picture, both effects should be taken into account using the same framework, since they will be governed by the same set of physical parameters (ALP mass and coupling constant). In the case of ALP masses  $m_a>>10^{-10}$ eV, the energies at which the photon/axion oscillation occur in the IGMF are $>>$ 1 TeV and thus not detectable with current gamma-ray instruments. In those cases the photon/axion oscillation in the source would be the only effect that could potentially be detected.

\item The photon/axion oscillation in the source (and its vicinity) can produce photon-flux attenuations up to 30\%, as previously stated in the literature \cite{hooper,simet}. However, when using available models for gamma-ray emitting blob regions to set values of the {\bf B} field strength and the size of the region where the conversion can take place (we took a radius 10 times the size of the blob), we obtain photon-flux attenuations that are significantly lower.

\item The photon/axion oscillation in the IGM produces a photon-flux attenuation up to 30\% below the energies at which the EBL is important (but above $E_{crit}$ for the oscillation to be efficient). If the source redshift is larger than $\sim$0.1, this drop in intensity should be about 30\% and it shows up in all sources at the same energy. Hence, it presents relatively easy signature of the presence of ALPs. The Fermi-LAT instrument is expected to play a very important role in this search, since it is expected to detect thousands of AGN sources located at various redshifts (up to z$=$5), and at energies where the EBL is not relevant. The detection of such a photon intensity drop would set the value for the product $m_a^2 \cdot M$, under the assumption of a given IGMF strength. If such an intensity drop is not seen in the spectra, lower limits could be set.

\item Above energies at which the absorption of gamma-rays in the EBL become important, the photon/axion oscillation in the IGMF could produce both attenuation and enhancement in the photon flux, depending on the source distance and energy under consideration.

\item We find that decreasing the intensity of the IGMF strength does not necessarily decreases the photon-flux enhancements (axion boost factors). For a source located at z$=$0.5, B$=$0.1~nG produces higher photon-flux enhancements that B$=$1~nG. This result is somewhat unexpected since stronger {\bf B} fields allow for a more efficient photon/axion mixing. The reason for this result is the strong attenuation due to the EBL. If the photon/axion mixing is efficient, then many ALPs convert to photons which soon disappear due to the EBL absorption. Consequently, if the source distance is large, it ends up having a very small number of photons arriving at the Earth. On the other hand, if the photon/axion mixing is not that efficient (lower B field), then there is a higher number of ALPs traveling (towards the Earth), which act as a potential reservoir of photons. The net balance between the two processes is sensitive to the source distance, the energy considered and the EBL intensity. Given those parameters, there is always a {\bf B} value that maximizes the photon flux enhancements.
\end{itemize}

We have shown that the signatures of photon/axion oscillations may be observationally detectable with current gamma-ray instruments (Fermi/LAT and IACTs). Since photon/axion mixings in both the source and the IGM are expected to be at work over several decades in energy, it is clear that a meticulous search for ALPs in the (sub)GeV-(multi)TeV regime will be greatly enhanced by means of a joint effort of Fermi and current IACTs.

The main challenge in such detection comes from the lack of knowledge in conventional physics; namely the intrinsic source spectrum and EBL density and the intensity and configuration of the intergalactic magnetic field. In other words, the effect of the photon/axion oscillations could be attributed to conventional physics in the particular source and/or propagation of the gamma-rays towards the Earth. However, we believe that such photon/axion oscillations could be studied using several distant AGNs located at different redshifts, as well as the same distant AGN detected at distinct activity levels. The signatures of such effect being attenuations (at relatively low energies) and/or enhancements (at the highest energies) in the photon fluxes, that could be visible in the residuals from the ``Best-Model-Fit'' and the observational data.  

Recent work, like the potential detection of TeV photons from very distant (z $\sim$ 0.4) sources, or those ones reporting energy spectral indices being harder than 1.5 for relatively distant (z=0.1-0.2) AGNs, already pose substantial challenges to the conventional interpretation of the observed gamma-ray data. Both effects could be explained by oscillations of photons (using $\sim$0.1~nG for the IGMF strength) into light ALPs ($m_a\leq ~10^{-10}$ eV) with a photon/axion coupling constant close to current upper limits ($M_{11} \sim$ 0.114).

\begin{acknowledgments}
We thank P. Serpico for useful discussions and comments. We also acknowledge the anonymous referee for useful comments that allowed us to improve the manuscript. M.A.S.C. is very grateful for the hospitality of the SLAC during his visit, where most of this work was done. This work was supported by the Spanish I3P-CSIC and AYA2005-07789 grants, by the SLAC/DOE Contract DE-AC02-76-SF00515, the Spanish Ministerio de Educaci\'on y Ciencia and the European regional development fund (FEDER) under project FIS-2008-04189, and Junta de Andaluc\'ia under project P07-FQM-02894.
\end{acknowledgments}

\end{document}